\documentclass[a4paper,fleqn,usenatbib]{mnras}

\usepackage{newtxtext,newtxmath}

\usepackage[T1]{fontenc}
\usepackage{ae,aecompl}

\usepackage{graphicx}	
\usepackage{mathrsfs}
\usepackage{amssymb}	
\usepackage{array}

\newcommand{\Msol}{\mbox{$M_{\odot}$}}
\newcommand{\Lsol}{\mbox{$L_{\odot}$}}
\newcommand{\Halpha}{\mbox{H$\alpha$} }

\title[CO(5-4) emission in a typical clumpy galaxy]{\LARGE ALMA constraints on star-forming gas in a prototypical $z$\,=\,1.5 clumpy galaxy: the dearth of CO$(5$$-$$4)$ emission from UV-bright clumps}

\author[A. Cibinel et al.]{A. Cibinel,$^{1,2}$\thanks{A.Cibinel@sussex.ac.uk}
E. Daddi,$^{2}$
F. Bournaud,$^{2}$
M. T. Sargent,$^{1}$
E. le Floc'h,$^{2}$
G. E. Magdis,$^{3,4}$
\newauthor
M. Pannella,$^{5}$
W. Rujopakarn,$^{6,7}$
S. Juneau,$^{2}$
A. Zanella,$^{8}$
P.-A. Duc,$^{2}$
P. A. Oesch,$^{9}$
\newauthor
D. Elbaz,$^{2}$
P. Jagannathan,$^{10}$
K. Nyland,$^{11}$
and T. Wang$^{2}$
\\
$^{1}$Astronomy Centre, Department of Physics and Astronomy, University of Sussex, Brighton, BN1 9QH, UK\\
$^{2}$CEA Saclay, DSM/Irfu/Service d'Astrophysique, Orme des Merisiers, F-91191 Gif-sur-Yvette Cedex, France\\
$^{3}$Dark Cosmology Centre, Niels Bohr Institute, University of Copenhagen, Juliane Mariesvej 30, DK-2100 Copenhagen, Denmark \\
$^{4}$Institute for Astronomy, Astrophysics, Space Applications and Remote Sensing, National Observatory of Athens, GR-15236 Athens, Greece\\
$^{5}$Faculty of Physics, Ludwig-Maximilians University, Scheinerstr. 1, 81679 Munich, Germany \\
$^{6}$Kavli Institute for the Physics and Mathematics of the Universe (WPI), The University of Tokyo Institutes for Advanced Study, \\
The University of Tokyo, Kashiwa, Chiba 277-8583 \\
$^{7}$Department of Physics, Faculty of Science, Chulalongkorn University, 254 Phayathai Road, Pathumwan, Bangkok 10330, Thailand \\
$^{8}$European Southern Observatory, Karl Schwarzschild Stra{\ss}e 2, 85748 Garching, Germany \\
$^{9}$Geneva Observatory, Universit\'e de Gen\`eve, Chemin des Maillettes 51,1290 Versoix, Switzerland \\
$^{10}$National Radio Astronomy Observatory, 1003 Lopezville Road, Socorro, NM 87801, USA\\
$^{11}$National Radio Astronomy Observatory, Charlottesville, VA 22903, USA
}

\date{Accepted XXX. Received YYY; in original form ZZZ}

\pubyear{2017}

\begin{document}
\label{firstpage}
\pagerange{\pageref{firstpage}--\pageref{lastpage}}
\maketitle

\begin{abstract}
We present deep ALMA CO(5-4) observations of a main sequence, clumpy galaxy at $z$\,=\,1.5 in the HUDF. Thanks to the $\sim$0$\farcs$5\ resolution of the ALMA data, we can link stellar population properties to the CO(5-4) emission on scales of a few kiloparsec. We detect strong CO(5-4) emission from the nuclear region of the galaxy, consistent with the observed $L_{\rm IR}$--$L^{\prime}_{\rm CO(5-4)}$ correlation and indicating on-going nuclear star formation. The CO(5-4) gas component appears more concentrated than other star formation tracers or the dust distribution in this galaxy. We discuss possible implications of this difference in terms of star formation efficiency and mass build-up at the galaxy centre.

Conversely, we do not detect any CO(5-4) emission from the UV-bright clumps.
This might imply that clumps have a high star formation efficiency (although they do not display unusually high specific star formation rates) and are not entirely gas dominated, with gas fractions no larger than that of their host galaxy ($\sim$50\%).
Stellar feedback and disc instability torques funnelling gas towards the galaxy centre could contribute to the relatively low gas content.
Alternatively, clumps could fall in a more standard star formation efficiency regime if their actual star-formation rates are lower than generally assumed. We find that clump star-formation rates derived with several different, plausible methods can vary by up to an order of magnitude. The lowest estimates would be compatible with a CO(5-4) non-detection even for main-sequence like values of star formation efficiency and gas content.
\end{abstract}




\section{Introduction}\label{sec:intro}
Improvements in sensitivity of sub-millimetre (sub-mm) interferometers over the past decade have enabled studies of the molecular gas content at galactic scales to reach very high redshifts.
While the first measurements typically targeted highly luminous starbursts (SBs) or quasars \citep[e.g.][]{Frayer+1998,Greve+05,Riechers+06,Maiolino+07}, we are now able to build samples of normal, main sequence (MS) galaxies with \emph{integrated} measurements of their CO content out to redshift $\sim$2.5 \citep{Daddi+08,Daddi+10a,Genzel+10,Geach+11,Tacconi+10,Tacconi+13,Bauermeister+13,Daddi+15}.
Such studies have demonstrated that MS galaxies at z\,$\simeq$\,2 exhibit large molecular gas fractions \citep[$\gtrsim$50\% of the baryonic mass,][]{Daddi+10a,Tacconi+10}, which can drive disc fragmentation and generate the giant star-forming clumps  \citep{Noguchi99,Bournaud+07,Agertz+09,Dekel+09} that characterize the complex morphologies often observed at these redshifts  \citep{Elmegreen+07,Bournaud+08,ForsterSchreiber+09,Guo+12}. 
Conversely, \emph{resolved} studies of the molecular gas at high redshift are still in their infancy and, although there have been few attempts to probe normal galaxies \citep[e.g.][]{Genzel+13,Aravena+14},  current works have typically focused on the SB tail of the galaxy population due to observational challenges \citep{Carilli+10,Swinbank+11,ALMAVlahakis+15,Hatsukade+15,Thomson+15}. 

Molecular gas observations of clumpy, MS galaxies at scales comparable to the giant star-forming clumps (stellar masses of $10^{8-9}$M$_{\odot}$ and sizes of 0.1-1kpc), in particular, could offer important clues on the physical processes governing galaxy evolution.
Migration of the clumps towards the galaxy centre and the torques they generate could play an important role in regulating bulge formation, black hole feeding and quenching of star formation in high redshift galaxies \citep{Dekel+09,Martig+09,Gabor+13,Zolotov+15}. Clump lifetimes and consequently the impact of clump-driven dynamical processes are however determined by the interplay of stellar feedback, gas accretion from the surrounding discs and efficiency of star formation \citep{Krumholz_Dekel2010,Bournaud+14,Mandelker+17}. 
Strong radiative feedback and stellar winds could disrupt the clumps before they can reach the galaxy centre and significantly contribute to black hole or bulge growth \citep[e.g.][]{Hopkins+12,Genel+12}.
Powerful outflows are observed in high-$z$ clumps \citep{Genzel+11,Newman+12}, but theoretical works have shown that clumps can survive these outflows if they are able to re-accreate gas from the surrounding disc \citep{Bournaud+14}.

Our observational knowledge of the molecular gas reservoirs in high-$z$ clumps and how these compare with their host galaxies is, however, still limited. 
In particular, it is not known whether the star formation efficiency in these giant clumps is similar to that seen in normal galaxies falling on the Schmidt-Kennicutt relation \citep{Schmidt_1959,Kennicutt_1998b}, or instead they display shorter time-scales for molecular gas consumption.
Indirect evidence presented in  \citet{Zanella+15} for a $<$\,10\,Myr old clump suggests a high star-formation efficiency, similar to that observed in SB galaxies/SB nuclei, during the early stages of clump formation. Other studies find depletion time-scales of the order of $\sim2$\,Gyr for unresolved ensembles of clumps \citep{Freundlich+13}, i.e., similar efficiencies as those observed in normal local galaxies and galactic regions \citep{Bigiel+08,Bigiel+11}. More observations are therefore necessary to discriminate among different scenarios. 

Among the tracers of the different phases of molecular hydrogen ($H_{2}$), the CO(5-4) transition is particularly advantageous for studying star formation occurring in a dense, clumpy medium.
Excitation of the high-$J$  CO transitions requires $H_{2}$ number densities of order 10$^{4-5}$\,cm$^{-3}$ \citep{Solomon_VandenBout2005,Carilli_Walter2013,Greve+14}, which are much more similar to the densities reached in star-forming cores of molecular clouds than the critical densities of lower $J$ transitions \citep{BerginTafalla2007}.  
A single, linear correlation between line luminosity, $L^{\prime}_{\rm CO(5-4)}$, and total infrared (8-1000$\mu$m) luminosity, $L_{IR}$, holds for normal and SB galaxies both locally and at high redshift \citep[e.g.,][]{Bayet+09,Greve+14,Daddi+15,Liu+15}. This has been interpreted  as evidence that the CO(5-4) line is a good tracer of the gas actually undergoing star-formation, irrespectively of the host galaxy properties.
Moreover, the numerical simulations of \citet{Bournaud+15} demonstrated that giant star-forming clumps could be responsible for highly excited CO spectral line energy distributions (SLEDs) as observed in MS, high-redshift galaxies, suggesting that this transition is indeed a good tracer of dense gas within the giant clumps.
Further evidence for this was presented by \citet{Daddi+15} who attempted a decomposition of the CO SLED of a high redshift galaxy into the emission from a diffuse component and emission originating from a clump-dominated region and found the latter to display higher excitation, with a SLED plausibly peaking at the CO(5-4) transition.  
Finally, the CO(5-4) line is much brighter and hence more easily observable at high redshift than other tracers that produce a linear correlation with star formation rate (SFR), e.g., HCN \citep{Gao_Solomon2004}.
  
In this paper we present $\sim$0$\farcs$5 Atacama Large Millimeter/submillimeter Array (ALMA) observations of the CO(5-4) transition in a z\,$\simeq$1.5 MS galaxy in the Hubble Ultra Deep Field (HUDF) area  \citep{Beckwith+06}. The target galaxy, UDF6462 \citep{Bournaud+08}, is a clumpy disc galaxy, enabling us to study the molecular content during this crucial phase for galaxy formation. 
With more than 4h integration time, our observations reach a continuum sensitivity that is roughly a factor of 2-3 times deeper than other studies of the HUDF covering a similar wavelength range \citep[but extend over a much larger area, e.g., ][]{Dunlop+17,Walter+16}. 

The paper is organized as follows: in Section~\ref{sec:AncillaryData} we describe the properties and available data sets for the clumpy disc galaxy UDF6462. Clump identification and properties are discussed in Section~\ref{sec:clumpDescription}. Our new ALMA observations are presented in Section \ref{sec:ALMAdata} and the results obtained from these data discussed in Section~\ref{sec:Results}. We summarize our findings in Section~\ref{sec:Summary}.
Throughout this paper we use magnitudes in the AB system \citep{Oke_1974} and a Chabrier initial mass function \citep[IMF,][]{Chabrier_2003}.
When necessary, measurements for comparison samples are converted accordingly. Our definition of stellar mass correspond to the actual mass in stars, i.e., it is corrected for the fraction that is returned to the interstellar medium (ISM) by evolved stars and differs from the integral of the star formation history (SFH), sometimes used in the literature. 
We adopt a Planck 2015 cosmology ($\Omega_m$\,=\,0.31, $\Omega_\lambda$\,+\,$\Omega_m$\,=\,1 and $H_0$\,=\,67.7\,km\,s$^{-1}$\,Mpc$^{-1}$, \citealt{Planck+15}).


\section{UDF6462: Ancillary Data and Integrated Properties}\label{sec:AncillaryData}

For our analyses we use a number of ancillary multi wavelength datasets available for UDF6462, including \textit{Hubble Space Telescope} (\textit{HST}) imaging at  $\sim$\,kilo parsec resolution, infrared (IR) observations from the \textit{Spitzer} and \textit{Hersche}l satellites and radio maps obtained with the Karl G. Jansky Very Large Array (VLA).
We describe in detail these data in the following section. 
The main physical properties for UDF6462 are provided in Table \ref{tab:UDF6462prop}.

\begin{table}
	\centering
\caption{Physical properties of UDF6462. The rows are: (1) and (2) celestial coordinates. Coordinates marked with the $\dagger$ symbol are corrected for the astrometric offset discussed in Appendix \ref{app:AstrometryOff}; (3) optical spectroscopic redshift; (4) integrated stellar mass from 3D-HST  \citep{Skelton+14};  (5) total IR luminosity derived through fitting of the IR SED; (6) integrated SFR derived from the total IR luminosity and applying the \citet{Kennicutt_1998} $L_{\rm IR}$-SFR relation; (7) SFR estimated from the VLA 5cm (6GHz) observations of \citet{Rujopakarn+16}, (8) integrated SFR derived from the UV luminosity with no dust extinction correction; (9) CO(5-4) velocity integrated line intensity; (10) CO(5-4) line luminosity; (11) observed 1300\,$\mu$m continuum flux density from the ALMA observations; (12) integrated gas mass inferred from the IR SED fitting; (13) molecular gas mass (nuclear region) derived from the CO(5-4) line luminosity.}
          \begin{tabular}{m{6cm}  r} 
		\hline\hline 
	        RA (deg, J2000) & 53.18172  \\
	                                     &53.18176$^\dagger$ \\
	        Dec. (deg, J2000) & $-$27.78298\\
	                                     &  $-$27.78308$^\dagger$ \\
                $z_{\rm spec}$ & 1.570\\
                $\log(M_{\star}/\Msol)$ & 10.41$\pm$0.15  \\
                $L_{\rm IR}$\,($10^{10}\Lsol$) & 33$\pm$9  \\
                SFR$_{\rm IR}$\,($\Msol$/yr)  & 33$\pm$9 \\
                SFR$_{\rm radio}$\,($\Msol$/yr)  &  43$\pm$7 \\
                SFR$_{\rm UV}$\,($\Msol$/yr)  & 3.4$\pm$0.5\\
                $I_{\rm CO(5-4)}$\,(Jy\,km\,s$^{-1}$)  & 0.308$\pm$0.055 \\
                $L^{\prime}_{\rm CO(5-4)}$\,($10^9$ K\,km\,s$^{-1}$\,pc$^{-2}$)  & 1.65$\pm$0.39 \\
                $S_{\rm cont}$\,($\mu$Jy)  & 106.7$\pm$26.2 \\
                $M_{\rm gas}$\,($10^{10}\Msol$) &  2.0$^{+4.4}_{-1.4}$ \\
                $M_{\rm mol.}$\,($10^{10}\Msol$) & 1.9$\pm$1.6 \\
                		\hline 
	\end{tabular} \label{tab:UDF6462prop}
\end{table}

\subsection{Multiwavelength data}

In selecting our target we aimed to observe a typical, clumpy high redshift galaxy while maximizing the amount of ancillary data supporting the interpretation of the ALMA observations. 
UDF6462 is located within the HUDF area and hence has been observed by several spectroscopic and photometric surveys. 
Moreover, its morphological and stellar population properties have also been investigated by a number of other studies in the the literature \citep{Elmegreen_Elmegree2005,Bournaud+08,Guo+12}.

Spectroscopic redshifts for UDF6462 were measured from the 3D-HST grism data \citep{Momcheva+16}, from the FORS2 follow-up of the Great Observatories Origins Deep Survey South (GOODS-S) field described in \citet{Vanzella+2008} and from the SINFONI observations of \citet{Bournaud+08}. All measurements are consistent with a $z_{\rm spec}$\,=\,1.570.         

\begin{figure}
\begin{center}
\includegraphics[width=0.49\textwidth]{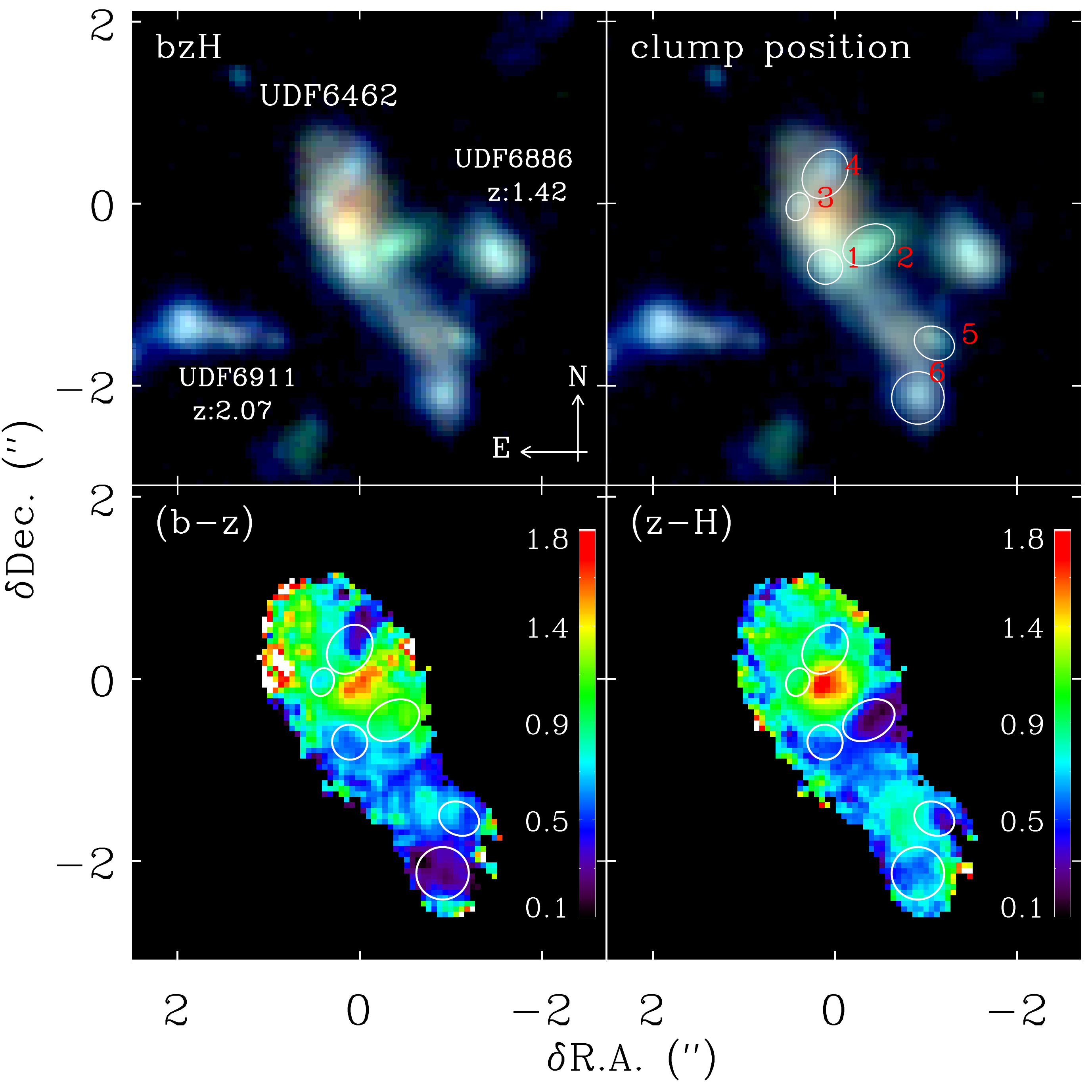}
\end{center}
\caption{\label{fig:UDF6462Stamps} From left to right and top to bottom: UDF6462 composite $b_{435}-z_{850}-H_{160}$ image from XDF without (left) or with (right) highlighting the position and size of the clumps identified in Section~\ref{sec:clumpDescription}  (white circles with red numbers); ($b_{435}-z_{850}$) colour map and ($z_{850}-H_{160}$) colour map. Two other galaxies at a different redshift are also indicated in the top left panel.
Axis coordinates are offsets relative to the phase centre of the ALMA observations.
}
\end{figure}

Extremely deep \textit{HST} observations  in the $UV_{275}$, $UV_{336}$, $B_{435}$, $V_{606}$, $i_{775}$, $i_{814}$, $z_{850}$, $Y_{105}$, $J_{125}$, $J_{140}$ and $H_{160}$ filters are available from the Ultraviolet Hubble Ultra Deep Field (UVUDF), HUDF and Cosmic Assembly Near-infrared Deep Extragalactic Legacy Survey (CANDELS) surveys and other follow-up campaigns \citep{Beckwith+06,Teplitz+13,Bouwens_et_al_2009,Ellis_et_al_2013,Grogin_et_al_2011,Koekemoer_et_al_2011}. In this paper, we specifically use $B_{435}$ to $H_{160}$ data extracted from the eXtreme Deep Field (XDF) obtained combining all of the available observations in this area and reaching a limiting magnitude of $\sim$30\,AB mag \citep{Illingworth+13}. The $UV_{275}$ and $UV_{336}$ images are extracted from the Hubble Deep UV (HDUV) data release (Oesch et al. 2017, submitted), which includes the v2 release of the UVUDF data base \citep{Rafelski+15}. 
The \textit{HST} stamp images and colour maps of UDF6462 are shown in Fig.~\ref{fig:UDF6462Stamps}.

Near-  to  far-infrared (FIR) coverage at wavelengths between 3.6$\mu$m and 500$\mu$m is also available from the \textit{Spitzer}, GOODS-Herschel and Herschel Multi-tiered Extragalactic Survey (HerMES) surveys  \citep{Dickinson+03,Elbaz+11,Oliver+2012}.  
At these long wavelengths source confusion becomes severe and flux deblending is necessary. 
Our mid-IR and FIR photometry of UDF6462 is based on new, improved deblending algorithms (see \citep{Liu+17}, for fluxes at $\lambda$\,$\leq$\,250\,$\mu$m and Wang et al., in preparation, at $\lambda$\,$>$\,250\,$\mu$m). 
At even longer wavelengths,  0$\farcs$3\,$\times$\,0$\farcs$6 resolution observations of the radio synchrotron emission were obtained from the VLA C-band (6GHz, $\sim$\,5\,cm) survey described in \citet{Rujopakarn+16}.

Finally, maps of $H\alpha$, [NII] and [SII] emission and the velocity field in UDF6462 were derived from the SINFONI integral-field spectroscopic observations presented in \citet{Bournaud+08}. 
Together with providing a kinematical classification of UDF6462, these data also enabled the derivation of gas-phase metallicities in different regions of the galaxy and thus provide important constraints on the CO-to-$H_2$ conversion factor ($\alpha_{\rm CO}$) in UDF6462 (see Section~\ref{sec:Mgas_SFE}).


\subsection{Integrated stellar mass and star formation rate estimates} \label{sec:SFRMassInt}

For the integrated stellar mass of UDF6462 we adopt the value of log(M$_{\star}$/$\Msol$)\,=\,10.4 published by the 3D-HST team and derived from the multiwavelength photometric catalogue (UV to NIR) of \citet{Skelton+14}. 
Independent spectral energy distribution (SED) fits that we performed using the \citet{Guo+13} CANDELS catalogue and different sets of stellar population models yielded consistent results within 0.15 dex \citep{Pannella+15}. In the following we will take this value to represent the typical uncertainty on the stellar mass of UDF6462.

\begin{figure}
\begin{center}
\includegraphics[width=0.48\textwidth]{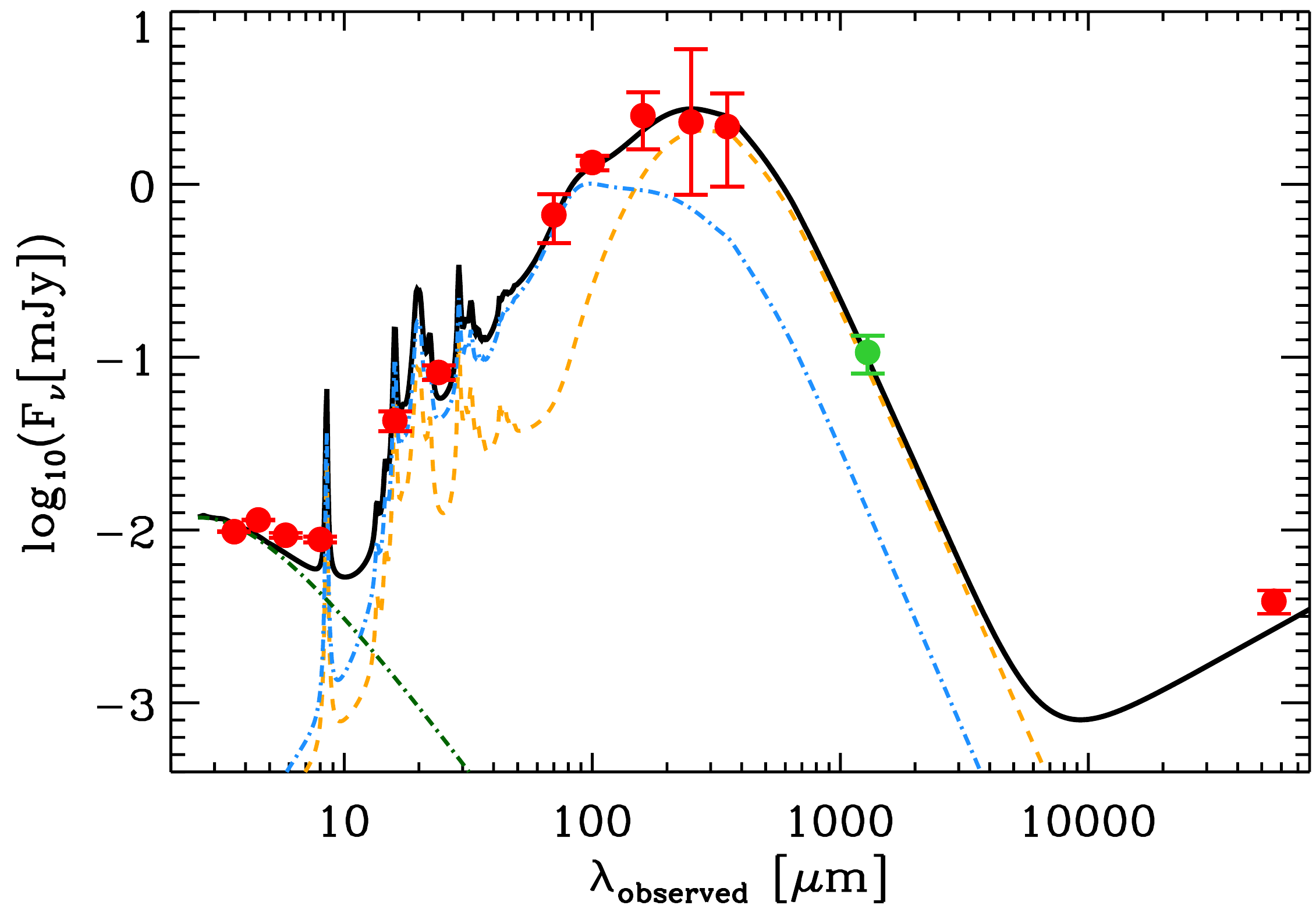}
\end{center}
\caption{\label{fig:IRSED} Observed IR/radio SED of UDF6462 (red circles with error bars) obtained combing the \textit{Spitzer}/\textit{Herschel} photometric data points at $\lambda$\,$\le$350$\mu$m  with the flux density measured from the ALMA continuum image at $\lambda \simeq1300\mu$m (Section~\ref{sec:Continuum}, shown as green point)  and the $\sim$5cm radio flux from the VLA observations of \citet{Rujopakarn+16}.
The thick black line shows the best-fitting \citet{Draine_Li_2007} model derived in Section~\ref{sec:SFRMassInt}, with the stellar contribution included (green dash-dotted line). The blue dashed-dotted and orange dashed lines are  the contributions from the diffuse ISM component and the photodissociation regions, respectively (see \citealt{Magdis+2011,Magdis+12} for a discussion on the fitting technique and components).}
\end{figure}

Thanks to the exquisite multiwavelength data sets available,  we can calculate the integrated SFR of UDF6462 using different approaches. 
Our fiducial estimate of the dust-obscured SFR is obtained by fitting the IR (3.6$\mu$m to 350$\mu$m) and radio photometry of UDF6462 with the \citet{Draine_Li_2007} models, following the approach described in \citet{Magdis+2011,Magdis+12}. 
In performing this fit, we included the continuum flux measurement at 1300$\mu m$ derived from the ALMA data presented in Section~\ref{sec:Continuum}, while we did not use the 500$\mu$m data point due to uncertain deblending for this galaxy even in our improved photometric catalogues.
The best-fitting IR-model  is shown in Fig.~\ref{fig:IRSED}.
It returns a total (8-1000\,$\mu$m) IR luminosity of $L_{\rm IR}$=\,(33$\pm$9)$\times10^{10}\Lsol$, corresponding to an SFR$_{\rm IR}$\,=\,(33\,$\pm\,$9)\,$\Msol$yr$^{-1}$ when using the \citet{Kennicutt_1998} L$_{\rm IR}$ to SFR conversion factor for a Chabrier IMF. 
Very similar results (within $\sim$\,0.1 dex) are obtained when fitting the SED using other commonly used codes \citep[e.g., \textsc{LePHARE} or \textsc{magpies},][]{Arnouts+99,daCunha+08} and different sets of IR templates \citep{Chary_Elbaz01,Dale_Helou_2002}.


From the HST data we can furthermore derive the unobscured component of the SFR. Using the $UV_{336}$ and $B_{435}$ band observations, we calculate a dust-uncorrected, rest frame UV luminosity of $L_{\rm UV} (1500\AA)$\,=\,(5.1\,$\pm$\,0.1)\,$\times$10$^{28}$\,ergs\,s$^{-1}$\,Hz$^{-1}$  and a SFR$_{\rm UV}$\,=\,(3.4\,$\pm$\,0.5)\,$\Msol$yr$^{-1}$ using the relation between UV luminosity and SFR in \citet{Daddi+04}, converted to a Chabrier IMF. 
Combining the IR and UV estimates, we obtain a total star formation rate of SFR$_{\rm tot}$\,=\,SFR$_{\rm IR}$\,+\,SFR$_{\rm UV}$\,=\,(37\,$\pm$\,9)\,$\Msol$yr$^{-1}$.

\begin{figure*}
\begin{center}
\includegraphics[width=0.7\textwidth,angle=90]{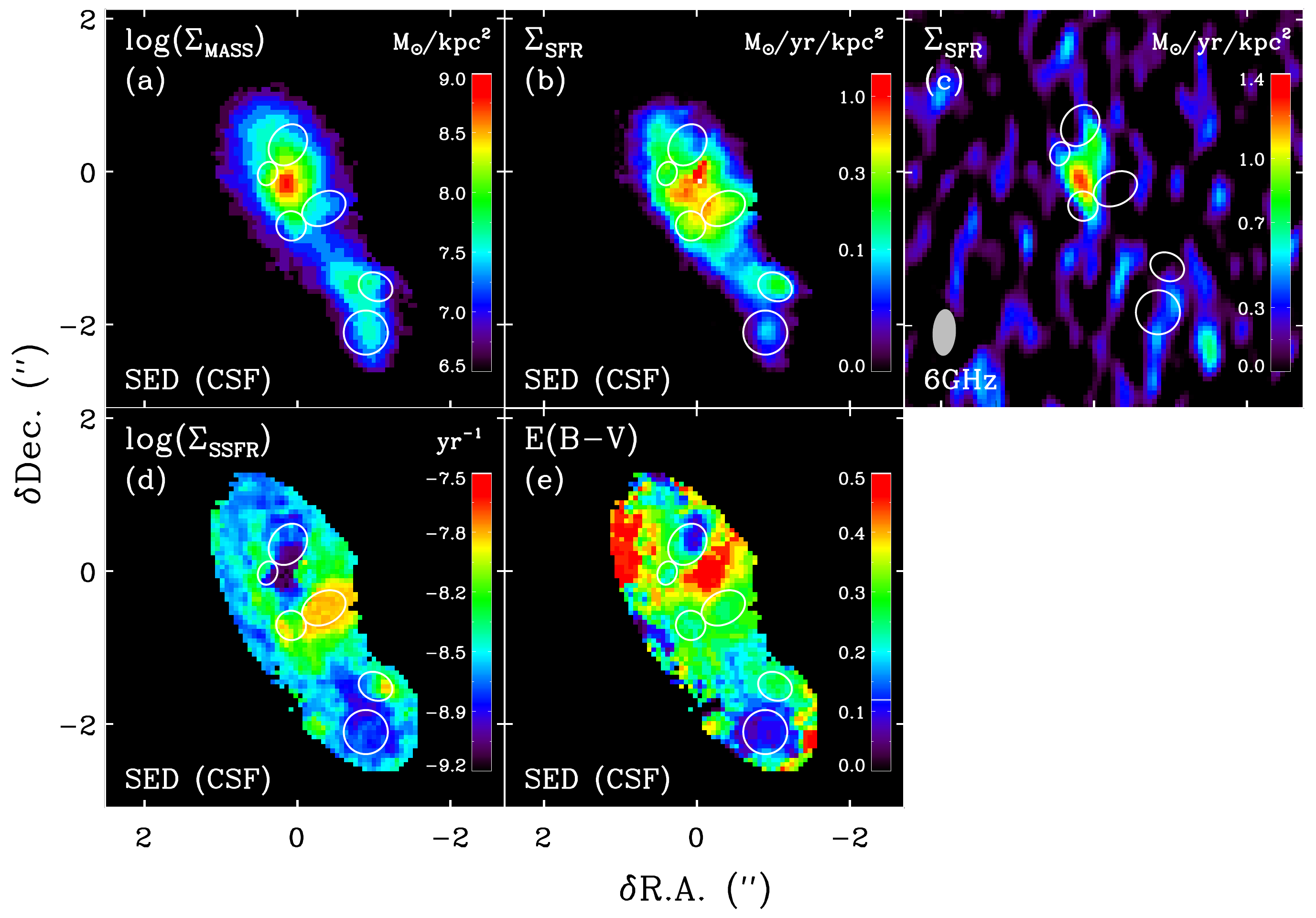}
\end{center}
\caption{\label{fig:UDF6462ResMaps} Resolved maps for UDF6462. From left to right and top to bottom: (a) map of stellar mass surface density;  (b) SFR surface density map;   (c) map of the SFR density obtained from the VLA 6GHz observations (VLA beam shown as grey ellipse in then bottom left corner);  (d) map of sSFR; (e) $E(B-V)$ dust extinction. All maps presented here, except panel (c), are derived from SED fitting with CSF history templates; maps obtained for a different set of template can be found in Figure \ref{fig:UDF6462ResMaps_Appendix}. Axis coordinates represent offsets with respect to the phase centre of the ALMA observations. White circles highlight the position of the clumps (see Figure \ref{fig:UDF6462Stamps}).
}
\end{figure*}

Finally, we can also derive an extinction-free SFR from the VLA C-band observations of \citet{Rujopakarn+16}. The radio map of UDF6462 is presented in the top right panel of Fig.~\ref{fig:UDF6462ResMaps}, showing clear radio emission from the galaxy central region.
The  6GHz flux density of UDF6462 was measured following the Gaussian fitting approach with the \textsc{PyBDSM} software described in \citet{Rujopakarn+16}. 
 To recover extended emission, the fitting was performed on several tapering scales and the one maximizing the signal-to-noise ratio (SNR) was chosen as the optimal one for source extraction.
With this method we obtained a total flux density of $S_{\rm 6GHz}$\,=\,(3.9\,$\pm$\,0.6)\,$\mu$Jy with a SNR=6.6 and extended over a Gaussian with beam-deconvolved full width at half-maximum (FWHM) of (0$\farcs$85\,$\pm$\,0$\farcs$23)\,$\times$\,(0$\farcs$42\,$\pm$\,0$\farcs$09). 
This flux density  corresponds to a rest-frame luminosity at 1.4GHz of 
\begin{eqnarray}
L_{\rm1.4GHz}&=&\frac{4\pi D_L}{(1+z)^{1-\alpha}}S_{\rm 1.4GHz}=\frac{4\pi D_L}{(1+z)^{1-\alpha}}S_{\rm 6GHz} \left(\frac{6}{1.4}\right)^{\alpha}\\
&=&(1.35\pm0.21)\times10^{23}{\rm W\,Hz}^{-1}\, ,
\end{eqnarray}
 when assuming a radio power law $S_{\nu}$\,$\propto$\,$\nu^{-\alpha}$ with $\alpha$\,=\,0.7 \citep{Condon_1992}. 
We then applied the radio luminosity to SFR conversion in \citet{Bell2003}, which for a Chabrier IMF implies an SFR$_{\rm radio}$\,=\,3.18$\times 10^{-22}$L$_{1.4GHz}$\,=\,(43\,$\pm$\,7)$\Msol$yr$^{-1}$.\\
The C-band flux corresponds to a logarithmic IR to radio flux ratio $q_{\rm IR}$\,=\,$\log\left(\frac{L_{IR}}{3.75\times10^{12} {\rm W}}\right)$\,--\,$\log\left(\frac{L_{1.4GHz}}{{\rm W\,Hz^{-1}}}\right)$\,=\,2.40$\pm$0.16, i.e., UDF6264 lies within the scatter of the observed local IR-radio correlation \citep[$q_{\rm IR}$\,$\simeq$2.6$\pm$0.3, e.g.,][]{Yun+01,Bell2003}.

To summarize, SFR values derived from  IR+UV and radio data are consistent within their uncertainties. For the reminder of the paper, we will thus use the average of the two measurements, $<$SFR$>$\,=\,(40\,$\pm$\,5)$\Msol$yr$^{-1}$, as our fiducial value for the SFR of UDF6462. 
With a stellar mass of $\log(M_{\star}/\Msol)$\,=\,10.4, UDF6462 is therefore slightly below  (0.3\,dex) the characteristic mass $M^{*}$ of the mass function of star-forming galaxies \citep{Ilbert+2013} and lies on the $z=1.5$ MS ($\Delta_{\rm MS}$\,=\,SFR/\,SFR$_{\rm MS}$\,=\,1.2, see also Fig.~\ref{fig:MSClumps}).

\subsection{Quantifying the AGN contribution}\label{sec:AGNcontribution}

\begin{figure*}
\begin{center}
\includegraphics[width=0.85\textwidth]{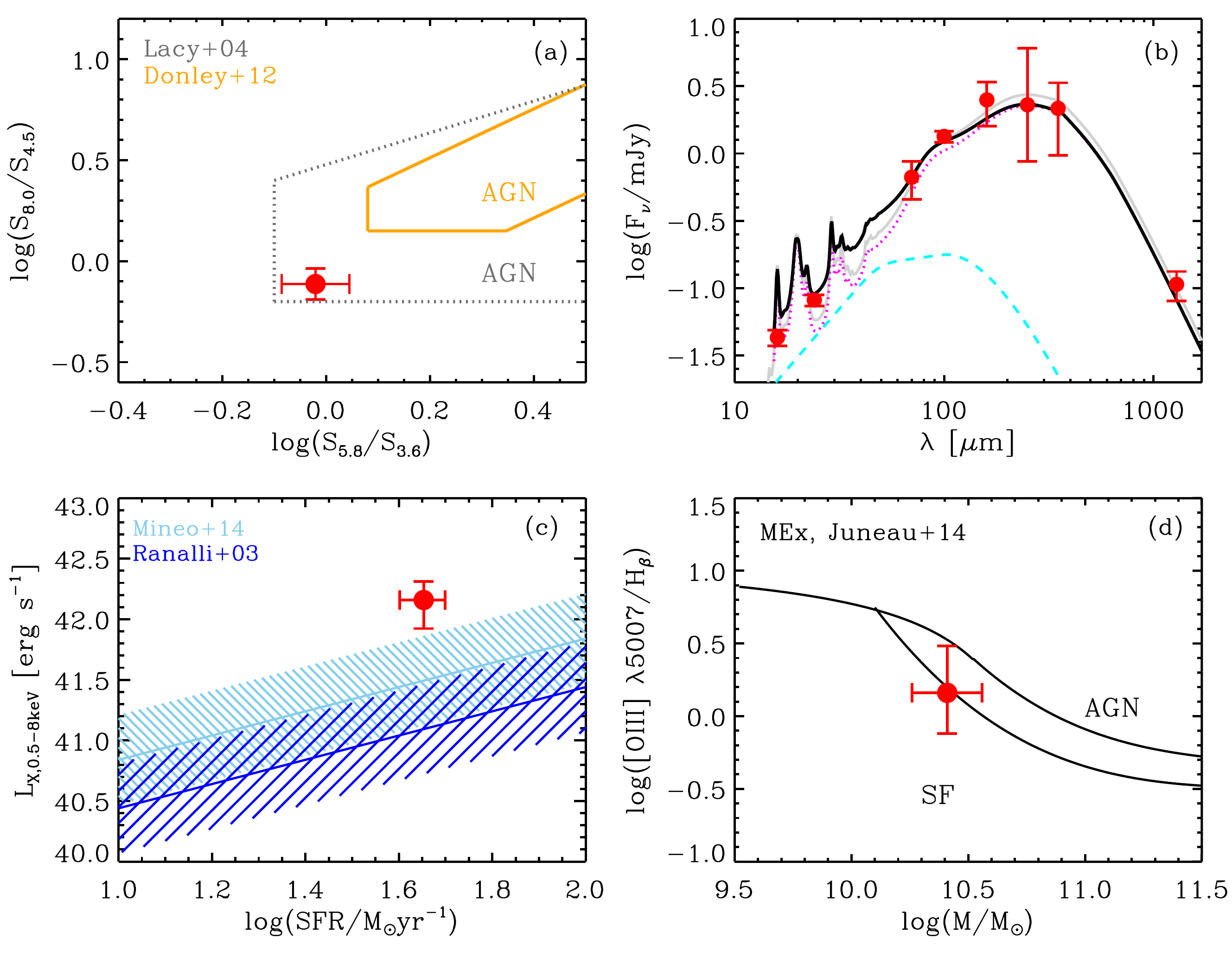}  
\caption{Diagnostic plots quantifying the presence of an AGN in UDF6462. 
(a): Position in the IRAC colour-colour plane. The dotted grey line corresponds to the original locus of AGNs defined in \citet{Lacy+04} and the solid orange line is the refined criterion minimizing contamination from star-forming galaxies described in \citet{Donley+12}.    
(b): Decomposition of the IR SED of UDF6462 into a purely star-forming (doted, magenta line) and AGN (dashed, cyan line) component (sum of components shown as thick black line). The AGN component is modelled with the template of \citet{Mullaney+11}; for the star-forming host we have employed the best-fitting SED obtained in Section \ref{sec:SFRMassInt} (see  Fig.~\ref{fig:IRSED}. Best-fit without AGN shown again here as grey line).
(c): Position of UDF6462 in the X-ray luminosity versus SFR plane. The shaded light blue and dark blue areas highlight the average relations and scatter for purely star-forming galaxies derived in \citet{Mineo+14} and \citet{Ranalli+03}, respectively (for the latter relation, we use the parametrization in \citet{Mineo+14} accounting for the conversion of $L_{\rm X,0.2-10keV}$ into $L_{\rm X,0.5-8keV}$). Both relations have been converted to a Chabrier IMF. The $L_{\rm X,0.5-8keV}$ X-ray luminosity of UDF6462 is taken from the 4\,Ms catalogue of  \citet{Xue+11}.
(d): Position of UDF6462 in the \textsc{MEx} diagnostic diagram of \citet{Juneau+14}.
}
\label{fig:AGNcont}
\end{center}
\end{figure*}

To understand whether UDF6462 hosts an active galactic nucleus (AGN) and whether its IR/radio and CO line emission could be significantly affected by it, we investigate several AGN diagnostics, which we present in Fig.~\ref{fig:AGNcont}.

As mentioned above, we estimated the rest-frame 1.4GHz luminosity of UDF6462 to be 1.36\,$\times10^{23}$W\,Hz$^{-1}$.
We note that this galaxy is not detected in the VLA 1.4GHz maps of GOODS-South published in \citet{Miller+2008}, placing a $3\sigma$ upper limit on the 1.4GHz luminosity(flux) of $2.46\times10^{23}$W\,Hz$^{-1}$ (19.5$\mu$Jy). 
Both estimates are below the threshold of 10$^{24}$W\,Hz$^{-1}$, above which powerful radio-loud AGNs typically dominate the radio luminosity function \citep[e.g.][]{Sadler+02,Smolcic+08}.

The IR properties of UDF6462 are also consistent with little contamination from an AGN component. This is shown in the upper two panels of Fig.~\ref{fig:AGNcont}. In panel (a) in particular, the IRAC [3.6]-[5.8] and [4.5]-[8.0] colours of UDF6462 place it in the region dominated by star-forming galaxies according to the criteria of  \citet{Donley+12} which was optimized for high-redshift galaxies. 
UDF6264 would be classified as an AGN according to the original cuts proposed by \citet{Lacy+04}, however it was shown in \citet{Donley+12} that, especially at high redshift, this selection can suffer from significant contamination from star-forming galaxies.
The above results are also confirmed by the decomposition of the IR SED of UDF6462 into an AGN and a star-forming component, shown in panel (b) of Fig.~\ref{fig:AGNcont}. 
To perform this decomposition, we used the AGN SED  template of \citet{Mullaney+11} together with the IR best-fitting template we obtained in Section~\ref{sec:SFRMassInt}  and the public \textsc{idl} code \textsf{decompir} for the fitting\footnote{https://sites.google.com/site/decompir/}. From this analysis,  we obtain an AGN contribution of $\sim$15\% to the total IR luminosity.
We note, however, that from the pure star-forming fit described in Section~\ref{sec:SFRMassInt} there is no evidence for the need of a second component. The percentage quoted above therefore likely represents an upper limit of the AGN contribution to the IR. 

X-ray observations of UDF6462 are available from the 4\,Ms and 7\,Ms Chandra Deep Field South data of \citet{Xue+11} and \citet{Luo+17}. These data provide X-ray luminosities equal to $L_{\rm X}$(0.5-8keV)\,=\,1.4$\times10^{42}$\,erg\,s$^{-1}$ and $L_{\rm X}$(0.5-7keV)\,=\,1.1$\times10^{42}$\,erg\,s$^{-1}$.  Although these place UDF6462 slightly above the locus of purely star-forming galaxies in the $L_{\rm X}$-SFR plane (see panel (c) of Fig.~\ref{fig:AGNcont}) the X-ray luminosity is well below the threshold of 3$\times10^{42}$\,erg\,s$^{-1}$ used to identify luminous AGN and UDF6462 is classified as a normal galaxy in both catalogues.

We can finally also use the stellar mass of this galaxy and the [OIII]/H$_{\beta}$ ratio measured from the 3D-HST grism spectrum \citep{Momcheva+16} to find the location of UDF6462 in the mass-excitation plane of \citet{Juneau+11,Juneau+14} (\textsc{MEx}\footnote{https://sites.google.com/site/agndiagnostics/home/mex}).  As discussed in \citet{Juneau+14}, at high redshift the dividing line between AGN and normal star-forming galaxies shifts towards higher stellar masses with respect to the local relation as a consequence of selection effects and evolution of the overall galaxy population. When calculating the AGN probability for UDF6264 using the \textsc{Mex} diagnostic, we therefore imposed a minimum line flux of 2.1$\times 10^{-17}$\,erg\,s$^{-1}$\,cm$^{-2}$\, corresponding to the detection limit of 3D-HST \citep{Momcheva+16} and allow for the evolution of the characteristic luminosity, $L^{*}$.
We find that UDF6462 lies below the limiting line dividing star-forming galaxies from AGN with a 64\% probability that the galaxy could be a purely SF system.

From the evidence above, we conclude that UDF6472 does not host a strong AGN and its radio and IR properties can largely be ascribed to pure star formation.


\section{Giant Clumps in UDF6462} \label{sec:clumpDescription}

The morphology of UDF6264 is characterized by several giant clumps, as shown in Fig.~\ref{fig:UDF6462Stamps}. 
As extensively discussed in \citet{Bournaud+08}, UDF6462 has a velocity gradient that is consistent with a rotationally supported disc (albeit disturbed by the presence of some kinematically distinct clumps at small scales). The complex morphology seen in the optical/NIR images of this galaxy is thus unlikely to be the outcome of a merger and can instead be explained with instability-driven formation of giant clumps, as observed in many z$>$1 galaxies.

\subsection{Resolved maps of physical parameters} \label{sec:ResolvedMaps}

Using the available \textit{HST} photometry spanning the wavelength range $\lambda_{\rm rest}$\,$\simeq$\,1000$-$6000\,$\AA$, we derived resolved colour maps and maps of stellar mass, SFR and dust extinction for UDF6462 following the approach in \citet{Cibinel+15}. The details of the generation of these maps can be found in \citet{Cibinel+15}.
Briefly, we performed pixel-by-pixel SED fitting to all available images, after homogenization of the SNR and convolution to match the $H_{160}$-band resolution. In this work the WFC3 $UV_{275}$ and $UV_{336}$ pass-bands were also included, providing better constraints on dust extinction and SFR.
   
In \citet{Cibinel+15} we used a set of \citet{Bruzual_Charlot_2003} templates with delayed, exponentially declining SFHs (delayed $\tau$-models), but for the purposes of this paper, we preferred to limit our library to templates with a constant star formation (CSF).  
It is known that SED fits in which the SFH and dust extinction are allowed to vary can result in degeneracies between highly obscured models and models dominated by an old stellar population. While this has a limited impact for stellar masses, it can lead to step-like spacial variations in the SFR maps, especially when individual pixels are fit independently from the others, as is the case in our analysis.
To minimize such degeneracies, we therefore decided to fix the SFH and consider the maps obtained with CSF templates as our fiducial  estimate.
Specifically, we used CSF models with ages between 100\,Myr and 2\,Gyr and a solar metallicity (in agreement with the estimate in 
\citealt{Bournaud+08}) and corrected for internal dust extinction assuming a \citet{Calzetti+00} law with $E(B-V)$ ranging between 0 and 0.9 mag.
We note that neither from the observed $(b_{435}-z_{850})$ and $(z_{850}-H_{160})$ colours nor applying the $UVJ$ colour criterion \citep{Williams+09} on the best-fitting SEDs is there evidence for a region with little current star formation in UDF6462, justifying our choice of a star-forming template.

\begin{figure}
\begin{center}
\includegraphics[width=0.46\textwidth,angle=90]{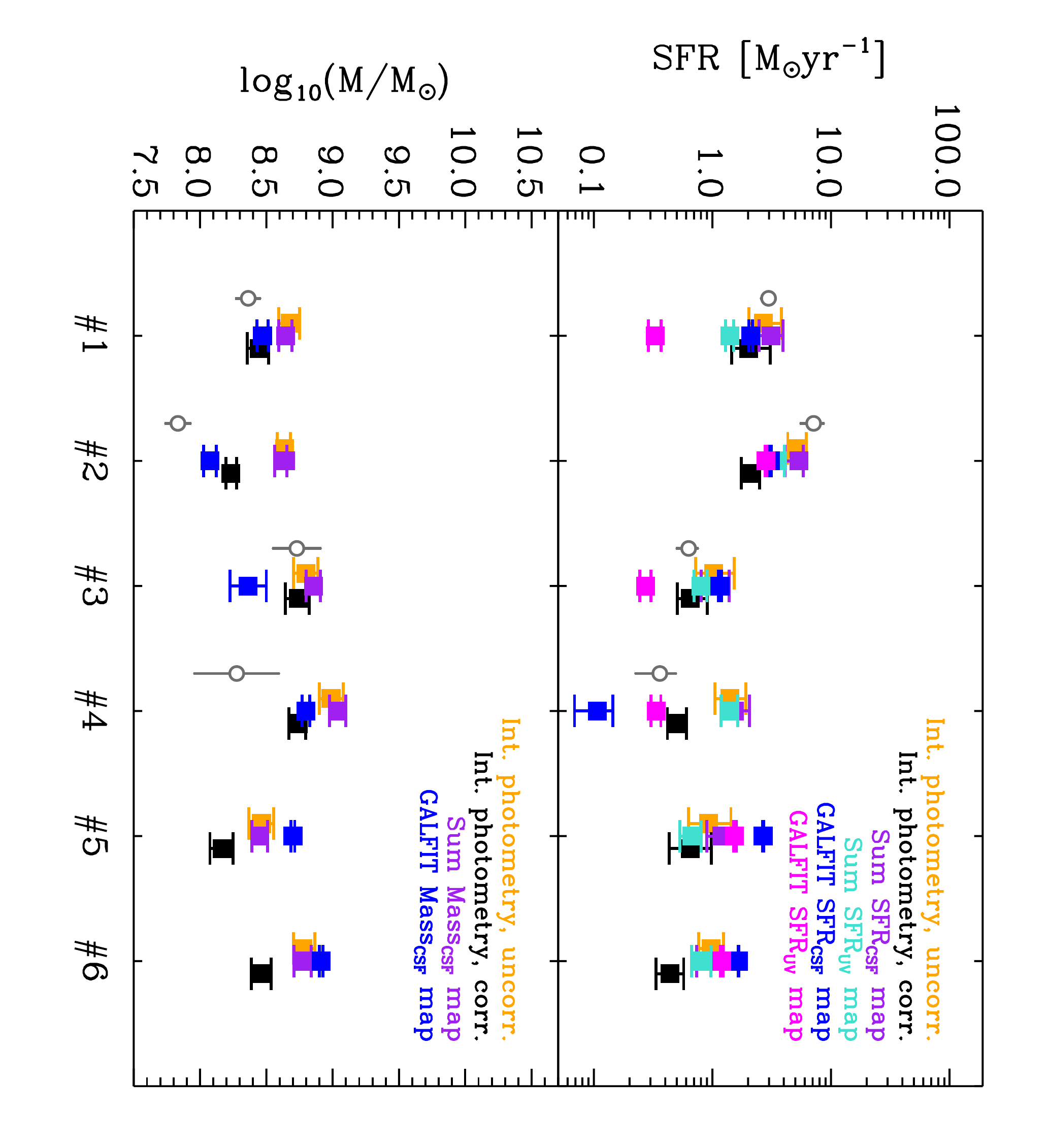}  
\caption{Clump SFRs (top panel) and masses (lower panel) calculated from:  SED fitting of the clump integrated photometry without background subtraction (\emph{orange}); SED fitting of the clump integrated photometry with background subtraction (\emph{black}); sum of the pixels in the resolved SFR or mass map obtained from SED-fitting with CSF models (\emph{purple}); sum of the pixels in the SFR map derived from the dust-corrected pixel UV luminosity (\emph{cyan}, SFR only); \textsc{galfit} decomposition (diffuse galaxy+clumps) of the mass or SFR map obtained from SED-fitting with CSF models  (\emph{blue}); \textsc{galfit} decomposition (diffuse galaxy+clumps) of the SFR map from the dust-corrected pixel UV luminosity   (\emph{magenta}, SFR only). See Appendix \ref{app:ClumpsProps} for details on the different methods.
Error bars correspond to the formal uncertainties in the derived parameters.
We show with empty grey circles the average SFR and masses derived in \citet{Guo+12} for the four clumps that we have in common with their sample. Note that, as discussed in Section~\ref{sec:ClumpsExtraction}, although we visually confirmed that these clumps are spatially coincident, it was not possible to verify if they also extend over exactly the same areas and hence whether their estimates are directly comparable with those presented here.}
\label{fig:ClumpsSFR}
\end{center}
\end{figure}

We also generated a less model dependent version of the SFR map by applying the $L_{\rm UV}$-SFR relation of \citet{Daddi+04}, where the pixel-based UV luminosities were calculated by interpolating the $UV_{336}$ and $B_{435}$ photometry and corrected for dust extinction using the  $E(B-V)$ versus $(b_{435}-z_{850})$ relation described in \citet{Daddi+04} and a \citet{Calzetti+00} extinction curve. 

For illustration, we show in Fig.~\ref{fig:UDF6462ResMaps}  the maps of stellar mass, SFR and specific SFR (sSFR) derived from the CSF SED fit, the other versions of these maps can be found in Fig.~\ref{fig:UDF6462ResMaps_Appendix} of Appendix \ref{app:ClumpMeasurements}.
When summing up the individual pixels in the resolved maps calculated via CSF SED fitting, we obtain a total stellar mass of 10$^{10.2}\Msol$ and an SFR of 40\,$\Msol$ yr$^{-1}$, which are in good agreement with the integrated properties discussed above. The SFR map derived from the corrected $L_{\rm UV}$ results in an SFR\,=\,26\,$\Msol$ yr$^{-1}$, suggesting that we are underestimating the dust extinction in this case. We further discuss dust extinction in the resolved maps in Section~\ref{sec:NuclearRegion} and Appendix \ref{app:ClumpMeasurements}.

\subsection{Clump identification}\label{sec:ClumpsExtraction}

To characterize the location of the clumps, we followed an approach similar to that outlined in \citet{Guo+12}, who studied clump properties for a sample of galaxies in the CANDELS field, also including UDF6462. 
Specifically, we identified as clumps those structures that are brighter than the underlying galactic diffuse component at the 5$\sigma$ level.
We characterized the diffuse component via both a non-parametric approach, i.e., by smoothing the galaxy images with a 4-pixel Gaussian kernel, and by modelling the galaxy surface brightness with \textsc{galfit} \citep{PengC2002,PengC2010}. 
In both cases we subtracted the diffuse component of the galaxy and run the Source Extractor code (\textsc{SExtractor}, \citealt{Bertin_Arnouts_1996}) on the residual image; sources of at least 5 pixels with a residual flux above the 5$\sigma$ level were identified as clumps.
We performed the detection in the $b_{435}$, $z_{850}$ and $H_{160}$ bands independently. 
We obtained consistent results for the three bands (see Fig.~\ref{fig:AppResiduals} in Appendix \ref{app:ClumpsProps}) and using the two smoothing methods.
For easier comparison with previous works, we use the $z_{850}$ band as the reference, but our findings do not depend on the exact clump detection technique that we employ.

We identified six main clumps in UDF6462 which we show in Figures \ref{fig:UDF6462Stamps} and \ref{fig:UDF6462ResMaps} as white ellipses with extents given by the \textsc{SExtractor} petrosian radii.
The four Northern clumps (Clump ID 1 to 4) coincide with those described in the work of \citet{Guo+12}\footnote{We note here that no detailed information on the exact coordinates and radii of the clumps is provided in the work of \citet{Guo+12}. Hence, although we visually verified that these clumps are coincident with theirs, we cannot confirm whether the clumps also extend over exactly the same areas.}.

\subsection{Clump properties}\label{sec:ClumpProp}

Table \ref{tab:UDF6462prop_clump} summarizes the clump main physical properties that we derived from the optical images and resolved maps.
Clump observed sizes are the \textsc{SExtractor}  petrosian radii calculated on the residual $b_{850}$-band image, after subtraction of the diffuse component. These radii enclose most of the clump fluxes in the optical images, as shown by the white circles in Fig.~\ref{fig:UDF6462Stamps}. We also derived intrinsic clumps sizes by fitting with \textsc{galfit} the residual $b_{850}$-band image with multiple Gaussian sources having centres fixed at the clump positions. Given its petrosian radius (see Table \ref{tab:UDF6462prop_clump}), we estimate that Clump \#3 is unresolved at the $H_{160}$-band resolution and hence treat it as a point source for the rest of the paper.

To estimate the stellar mass and SFRs of the clumps we followed a number of approaches, an in-depth discussion of these methods is given in Appendix \ref{app:ClumpMeasurements}. 
Briefly, the techniques we explored differ in terms of how the clump SED fitting is performed (integrated clump photometry versus pixel-by-pixel from resolved maps), how SFR is modelled/traced (UV luminosity versus CSF templates) and whether or not a contribution for a galactic diffuse component is subtracted.
As evident in Fig.~\ref{fig:ClumpsSFR}, the derivation of the clump physical parameters is affected by systematics that  depend on all these assumptions.
While these systematics are less important for stellar masses, which typically agree within a factor of 2-3 for each individual clump, SFR values can differ by a factor $\sim$\,10 in some of the clumps.
We adopt the average of the different measurements for each clump as our preferred estimate for its mass and SFR but, to stress that these can be subject to large uncertainties, we will show the full range of possible values as error bars in our plots.

\begin{figure}
\begin{center}
\includegraphics[width=0.5\textwidth]{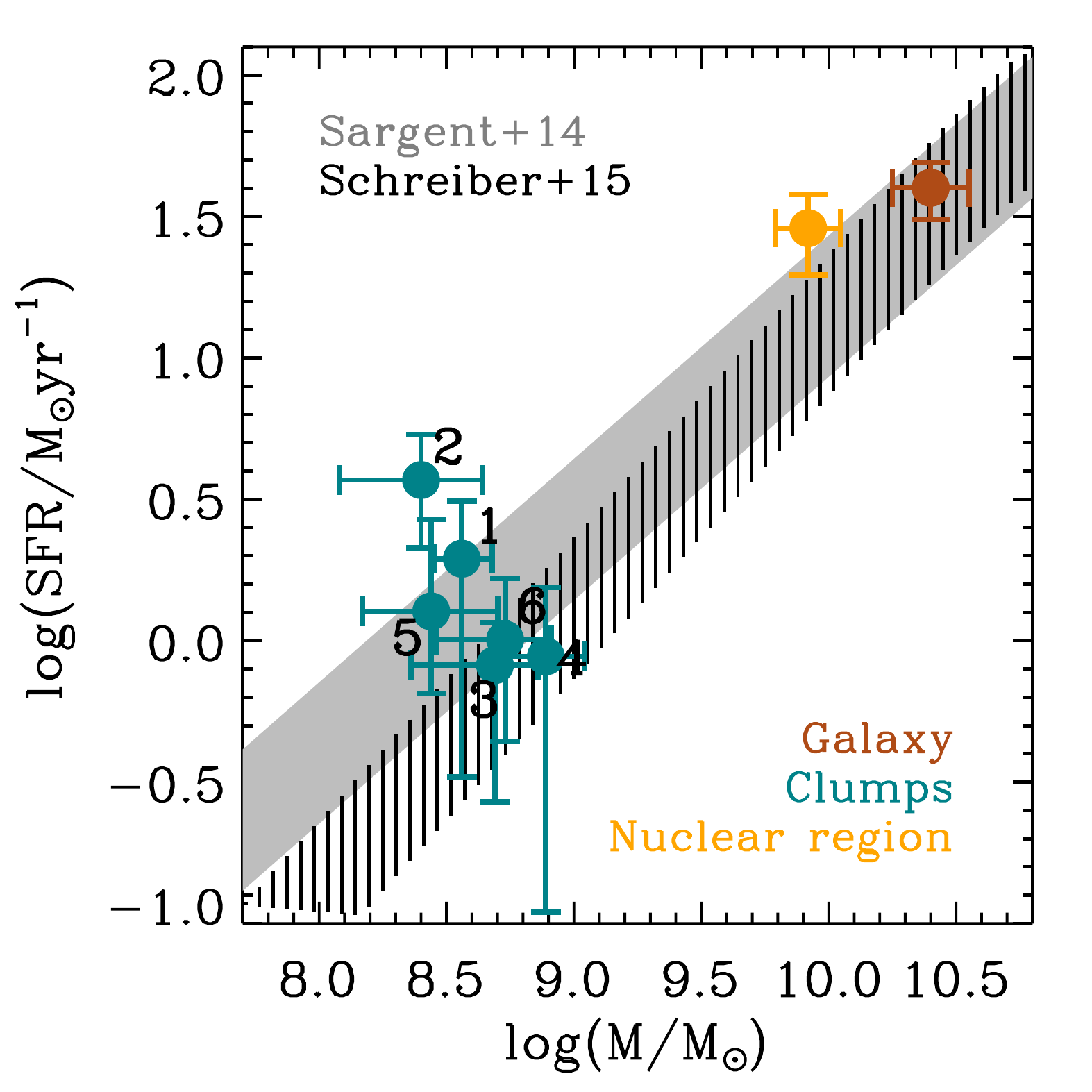}  
\caption{Position of UDF6462 and regions within it on the stellar mass versus star formation plane. Giant clumps are shown with blue circles annotated with numbers. The point highlighted with an orange symbol corresponds to the nuclear region and the brown one to the integrated parameters for the entire galaxy.  The shaded areas indicate the locus and typical scatter of the MS according to the parametrization of \citet{Schreiber+15} (dashed) and \citet{Sargent+14} (filled, light grey), extrapolated down to the clump masses (the original relations were derived for galaxy masses $>10^{9}\Msol$). 
 }\label{fig:MSClumps}
\end{center}
\end{figure}

The six clumps constitute on average $\sim$25\% and  $\sim$10\% of the total star formation and stellar mass in the galaxy, respectively.
Depending on the method applied, however, the contribution to the SFR varies between 10\% and 40\%  and between 6\% and 16\% for the stellar mass.
The position of the clumps in the SFR versus mass plane is shown in Fig.~\ref{fig:MSClumps}.
We show for reference the MS of star-forming galaxies as derived from galaxy integrated properties \citep[using the parametrization of the MS from][]{Sargent+14,Schreiber+15}. If this relation holds at the scale of clumps, our clump sample mostly consists of ``MS" clumps, with just one clump showing somewhat enhanced activity (\#2).

In terms of sSFR, clumps in UDF6462 have average values of the order of sSFR\,$\simeq$\,$2\times 10^{-9}$\,yr$^{-1}$, i.e., a stellar mass doubling time-scale of 1/sSFR$\simeq$\,$5\times 10^{8}$\,yr.
As mentioned above, these sSFR are comparable to those measured in MS star-forming galaxies at the redshift of UDF6462, however are lower than the sSFR that can be attained in the most massive giant molecular clouds, where a few percent of their gas masses ($\sim$\,$10^{5}\Msol$) is locked into stars with SFR of the order of $200-300 \times 10^{-6}\Msol$\,yr$^{-1}$ \citep[e.g.][]{Heiderman+10,Lada+10,Evans+14}, resulting in an approximate doubling time-scale of 1/sSFR$\simeq$\,$10^{7}$\,yr.
As visible in Fig.~\ref{fig:UDF6462ResMaps}, the clumps also have typically low to intermediate dust extinction.

\begin{table*}
	\centering
	\caption{Properties of clumps in UDF6462. Table columns are: (1) clump ID; (2) and (3) celestial coordinates -- coordinates marked with the $\dagger$ symbol are corrected for the astrometric offset discussed in Appendix \ref{app:AstrometryOff}; 4) the petrosian semi-major axis calculated using \textsc{SExtractor} in the $z_{850}$ band; (5) clump FWHM (major axis) from \textsc{galfit} Gaussian fit in the $z_{850}$ band (clump 3 is best modelled as a point source); (6) and (7) clump stellar mass and SFR (average of all available estimates described in Appendix \ref{app:ClumpMeasurements}); (8) colour excess from integrated clump photometry without diffuse component subtraction. The two numbers in the $E(B-V)$ column quote the values obtained from extinction maps constructed using either SED fitting with CSF models (first value) or the $E(B-V)$ versus $(b_{435}-z_{850})$ colour relation described in \citet{Daddi+04} and a \citet{Calzetti+00} extinction curve (second value); (9) and (10) $(b_{435}-z_{850})$ and $(z_{850}-H_{160})$ colours from integrated clump photometry without diffuse component subtraction; (11) upper limit on the CO(5-4) velocity integrated line flux,  $I_{\rm CO(5-4)}$,  derived assuming a clump line width of 100 km\,s$^{-1}$. 
} \label{tab:UDF6462prop_clump}
\begin{tabular}{ccccccccccc} 
\hline \hline 
      ID  & RA  & Dec. & SMA  & $\theta_{\rm major}$ & M$_{\star}$ & SFR  & $E(B-V)$ & $(b_{435}-z_{850})$  & $(z_{850}-H_{160})$  & $I_{\rm CO(5-4)}$   \\
          & (deg, J2000) & (deg, J2000) & ($^{\prime \prime}$) & ($^{\prime \prime}$)  &$\log(M/\Msol)$  & ($\Msol$/yr)  & (mag) & (mag) & (mag) &  (Jy\,km\,s$^{-1}$) \\ 
\hline \\
1        & 53.181717  & -27.783123   &  0.19  & 0.06 & 8.56$\pm0.12$       
                                              &   1.95$\pm$1.0     & 0.26/0.20      & 0.69$\pm$0.13  & 0.71$\pm$0.12   & $<$0.020  \\
                        & $^\dagger$53.181732 & -27.783199      \\             
  2        &  53.181583 & -27.783060 & 0.30  & 0.14 & 8.40$\pm0.29$         
                                                &   3.7$\pm$1.3              &  0.29/0.28     & 1.02$\pm$0.13  & 0.45$\pm$0.12 & $<$0.021\\
                         & $^\dagger$53.181599 & -27.783136 \\
 3        & 53.181812   & -27.782932 & 0.16  &  pt.\,s. &  8.69$\pm0.22$                    
                                                & 0.8$\pm$0.3   &  0.26/0.25     &  0.89$\pm$0.16 & 1.20$\pm$0.12 & $<$0.019 \\ 
                        & $^\dagger$53.181828 & -27.783009 \\                                            
 4        &  53.181728  & -27.782829 & 0.29 & 0.18 & 8.89$\pm0.14$              
                                                 & 0.9$\pm$0.6       & 0.18/0.19    &  0.67$\pm$0.13 &  1.01$\pm$0.12 & $<$0.021 \\
                        & $^\dagger$53.181744 & -27.782906 \\                                            
 5         & 53.181370  & -27.783344 & 0.23 & 0.25 &  8.44$\pm0.22$                        
                                                 & 1.3$\pm$0.8    & 0.25/0.19            & 0.69$\pm$0.16 &  0.75$\pm$0.16 & $<$0.021 \\
                        & $^\dagger$53.181385 & -27.783421 \\                                            
6        & 53.181408  & -27.783510  & 0.29  & 0.31 & 8.73$\pm0.19$                        
                                              &   1.0$\pm$0.4      & 0.11/0.11           &  0.32$\pm$0.13  & 0.74$\pm$0.13 & $<$0.025 \\
                        & $^\dagger$53.181423 & -27.783587 \\                                
		\hline
	\end{tabular} 
\end{table*}

\subsection{Nuclear region}\label{sec:NuclearRegion}

At the centre of UDF6462 there is a red nucleus, which displays a high SFR and very high dust extinction in the resolved maps (see Figures~\ref{fig:UDF6462Stamps} and \ref{fig:UDF6462ResMaps}). 
High attenuation towards the centre of galaxies is commonly found in high redshift galaxies with masses comparable to that of UDF6462 \citep[e.g.][]{Nelson+16}.
The significant radio emission over this area in the VLA radio map (see top right panel in Fig.~\ref{fig:UDF6462ResMaps}), as well as the results from the ALMA observations we describe in the next section, are further evidence for this region being truly star-forming. 

The red nucleus is also coincident with the peak of the mass distribution and is metal enriched with respect to the rest of the galaxy \citep{Bournaud+08}. 
 Moreover, the \textsc{galfit} model of the galaxy $H_{160}$-band surface brightness distribution yields a fairly concentrated profile with a S\'ersic index of $n$\,=\,2.5$\pm$0.1.
A \textsc{galfit} fit to the mass maps results in a similarly high index of $n$\,=\,3.0\,$\pm$0.1. 
The above properties are suggestive of a forming bulge, still hosting a non-negligible amount of (obscured) star formation (see also discussion in \citealt{Bournaud+08}).

\begin{table*}
\centering
\caption{Summary of ALMA observations for Project ID: 2013.1.01271.S. We provide a record of: date of observations, ALMA configuration, number of 12m antennas and maximum baseline in the given configuration, total on-source integration time, angular resolution (synthesized beam major axis, natural weighting),  average precipitable water vapour (PWV) during observations and maximum recoverable scale.}\label{tab:ObsLog}
      \begin{tabular}{cccccccc}		
      \hline \hline
Date & Configuration & Number of antennas & Max. baseline & Time on source &  Angular Resolution & PWV & Max. Scale\\
 & & & (km) & (hr) &  (arcsec) & (mm) & (arcsec) \\
 		\hline \\
2014 July 18 & C34-4/5 & 32 &  0.65 & 0.56 & 0.65& 0.58 & 8.4\\
2014 July 22 & C34-4/5 & 33 & 0.78 & 0.77 &  0.42 & 1.64 & 8.4 \\
2015 June 06 & C34-5  & 37 & 0.74 & 0.77 &  0.61 & 0.56 & 7.8 \\
2015 June 16 & C34-5 & 36 & 0.78 & 0.77 &  0.46 & 0.58 & 6.3 \\
2015 August 28 & C34-7/(6) & 40 & 1.6 & 0.77 &  0.26 &1.75 & 4.0 \\
2015 August 28 &C34-7/(6) & 40 & 1.6  & 0.77 &  0.20 &1.70 & 4.0 \\
\hline
	\end{tabular} 
\end{table*}

We note that in the work of \citet{Guo+12} a clump has been identified in this region (number 1, in their fig. 1). We also find some evidence for a clumpy structure (see e.g., Fig.~\ref{fig:AppResiduals})  that is slightly shifted eastward with respect to the centre of mass and the reddest peak in the ($b_{435}-z_{850}$) and ($z_{850}-H_{160}$) colour maps, although it is not as prominent at the other clumps.
It is possible that towards the centre of the galaxy a young clump in the foreground overlaps with the central mass concentration. 
However, given the complexity of the galaxy in this area, we do not attempt a separation of these two components and instead  we will consider the region as a whole and refer to it as the ``nuclear region" from now on. 

Due to the high dust extinction, SED-based SFR estimates for this region vary strongly depending on the assumed SFH and are consequently subject to large uncertainties. 
However, the high-resolution VLA observations, which are not affected by dust obscuration, enable us to derive more robust SFRs and also to quantify systematic biases in the SFR map from the SED fitting.
A comparison of the SFR surface density over the nuclear region as derived from the radio observations and the resolved maps is presented in Appendix \ref{app:ClumpMeasurements}. We find that the SED-based maps result in shallower SFR profiles and  underestimate the SFR in the nuclear region by roughly factors of 1.5 to 2.5 depending on the templates used.
For this reason, when considering the red nucleus of UDF6462, we use the radio image as our fiducial SFR tracer. 

As we discuss in more detail in the next section, the bulk of the CO(5-4) emission in our ALMA observations originates from this nuclear region. To be able to directly link its SFR and stellar mass to the molecular gas content, we use the best-fitting Gaussian parameters of the CO map (Section~\ref{sec:CO5-4Emission}) to define the extent of the nuclear region, instead of performing the extraction on the $z_{850}$ image as was done for the clumps.  
The stellar mass of the nuclear region is then obtained by summing up pixels in the mass map within an ellipse of major axis equal to the FWHM of the CO(5-4) Gaussian. The SFR is derived by integrating the radio best-fitting Gaussian over the width of the CO(5-4) Gaussian. We obtained $\log(M_{\star}/\Msol)$\,=\,9.92\,$\pm$\,0.13  and SFR\,=\,(29\,$\pm$\,9)\,$\Msol$ yr$^{-1}$ for the nuclear region (orange symbol in Fig.~\ref{fig:MSClumps}).


\section{ALMA Band-6 Follow-up} \label{sec:ALMAdata}

\subsection{Observations} \label{sec:ObsDetails}
ALMA Band 6 observations for UDF6462 were carried out during Cycle 2 with angular resolution ranging from 0$\farcs$2 to 0$\farcs$6 (Project ID: 2013.1.01271.S, PI: Cibinel). A log of these observations is provided in Table \ref{tab:ObsLog}.
We configured the first ALMA baseband which contains the redshifted CO(5-4) line -- observed and rest frequencies equal to $\nu_{\rm obs}$\,=\,224.23\,GHz and $\nu_{\rm rest}$\,=\,576.26\,GHz -- with a spectral resolution of 0.976\,MHz. The other three baseband were used in Time Division Mode (TDM) mode for continuum observations.
Considering line-free channels in all four bands, our continuum observations cover 7.4 GHz and the rest-frame wavelength range $\lambda_{\rm rest}$\,$\simeq$\,480-520\,$\mu$m (average observed wavelength $\lambda_{\rm obs}$\,$\simeq$\,1300\,$\mu$m).
We integrated for a total of 4.4hr on source, reaching sensitivities of $\sigma_{\rm cont}$\,=\,7.6$\mu$Jy/beam (natural weighting) for the continuum and  $\sigma_{\rm CO(5-4)}$\,=\,0.066mJy/beam over a 100km/s channel for the CO line.
According to the CO(5-4)-IR relation in \citealt{Daddi+15}, the $3\sigma$ line sensitivity corresponds to a SFR$\sim$3$\Msol$/yr$^{-1}$ in an individual resolution element.

Calibration of the ALMA raw data was carried out using Common Astronomy Software Applications \citep[\textsc{casa},][]{McMullin+07} and the standard ALMA pipeline.
Imaging of visibilities and the scientific analysis of the data cubes were instead mostly performed with the \textsc{gildas} software package \citep{Guilloteau_Lucas_2000}.
To maximize sensitivity, a natural weighting was used to generate the maps, resulting in a synthesized beam of FWHM $\theta_{\rm beam}$=0$\farcs$62$\times$0$\farcs$40.  The maximum recoverable scale of our observations is $\sim$4$^{\prime\prime}$. 

As we discuss in more detail in Appendix \ref{app:AstrometryOff}, there is an astrometric offset between the ALMA observations and \textit{HST} data products on the GOODS-S field that has to be taken into account when interpreting the data (see also \citealt{Dunlop+17, Rujopakarn+16,Barro+16}).
In this paper we only refer to \textit{HST} images which already incorporate the astrometric correction.
However, to allow an easy identification of the clumps/structures in the optical images and the comparison with previous studies, we also provide the uncorrected \textit{HST} celestial coordinates in Tables \ref{tab:UDF6462prop} and \ref{tab:UDF6462prop_clump}.

For reference and comparison of our data with previous studies, we also provide in Appendix \ref{app:AstrometryOff} the $\sim$\,1300$\mu$m integrated fluxes for four other galaxies detected in our continuum image. Three of these galaxies are in common with the sample of \citet{Dunlop+17} and \citet{Rujopakarn+16}.                              
 
\subsection{CO(5-4) line emission} \label{sec:CO5-4Emission}

\subsubsection{Nuclear Region}

\begin{figure*}
\begin{center}
\begin{tabular}{ccc}
\includegraphics[width=0.36\textwidth]{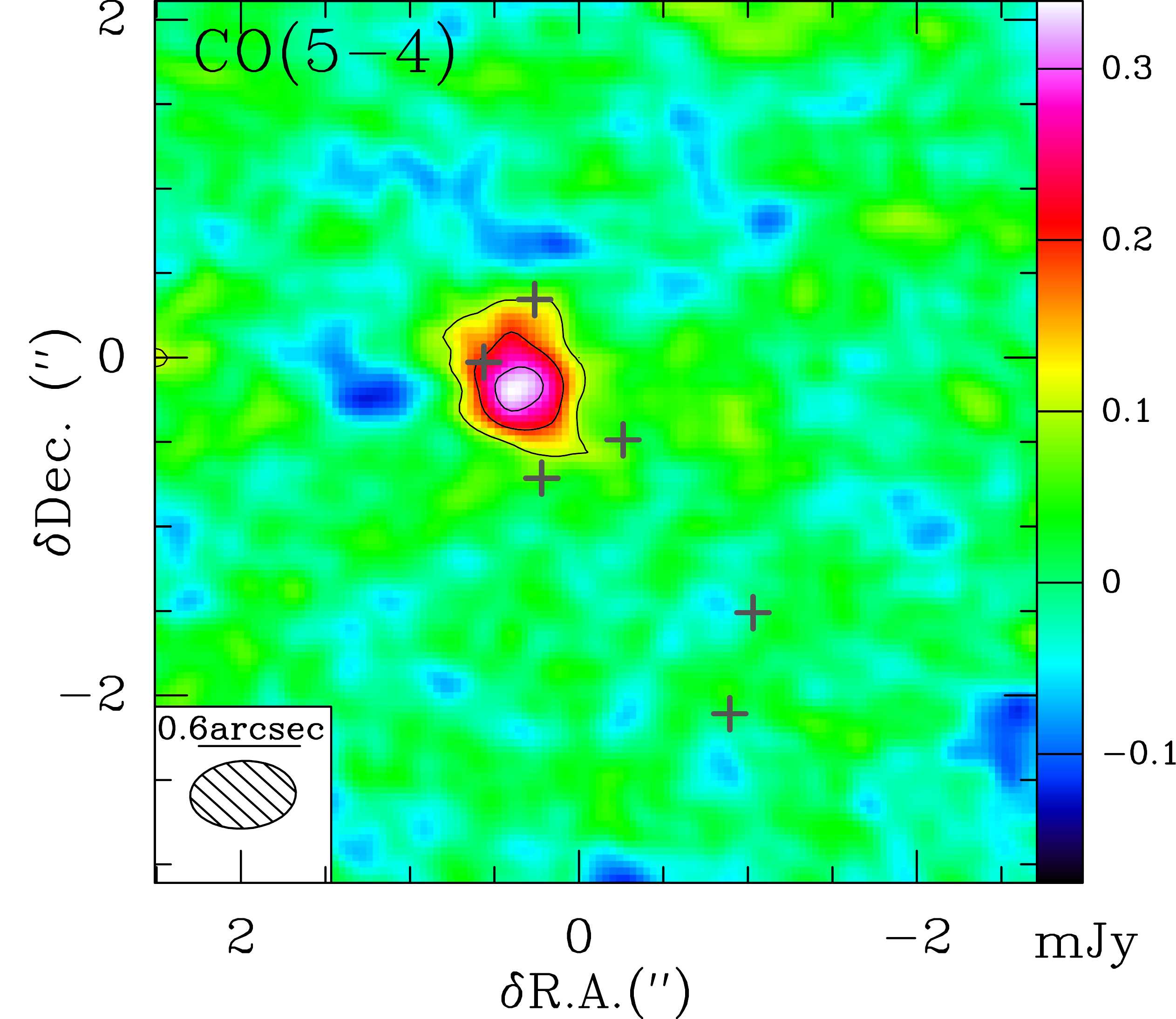}  
& \includegraphics[width=0.315\textwidth]{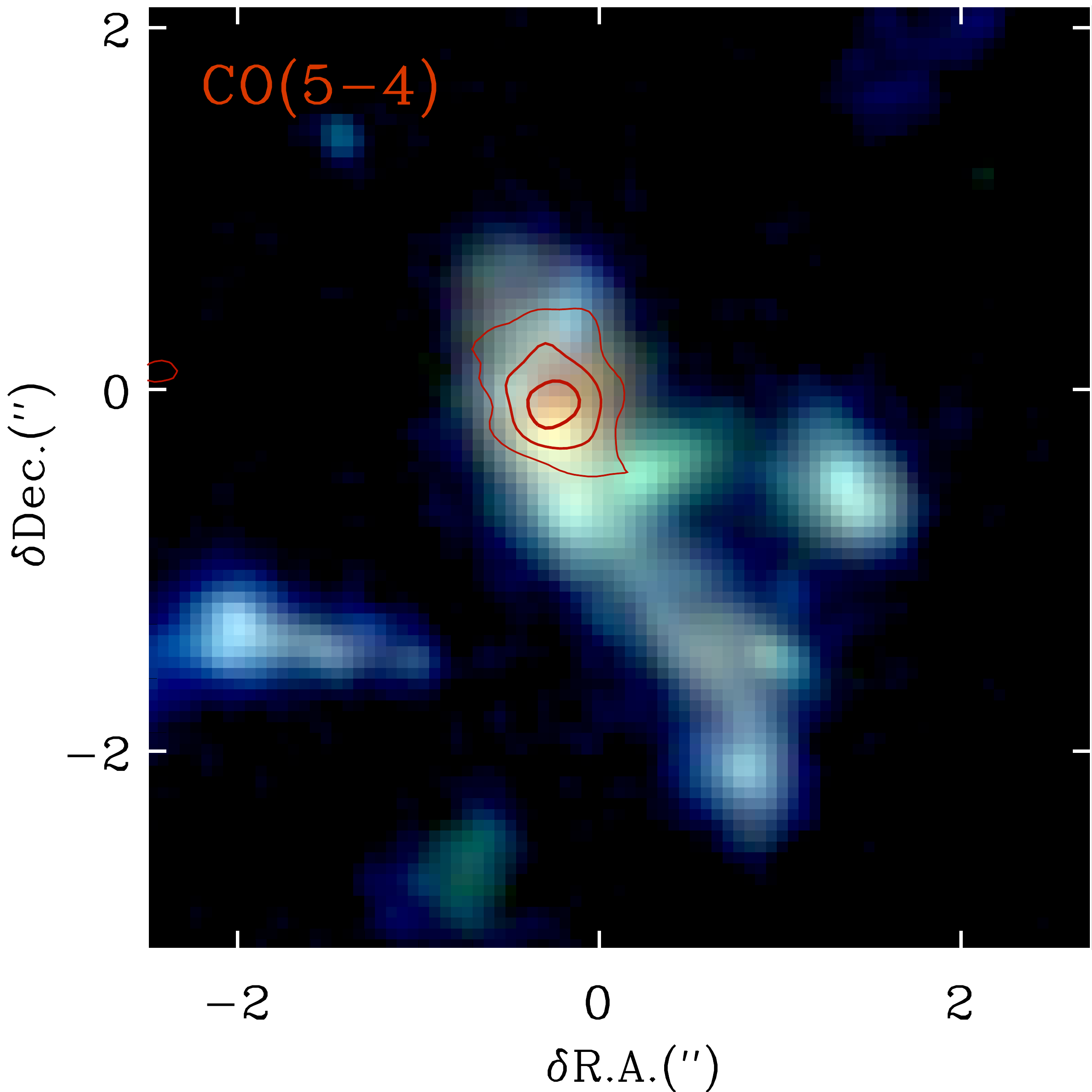} 
 & \includegraphics[width=0.315\textwidth]{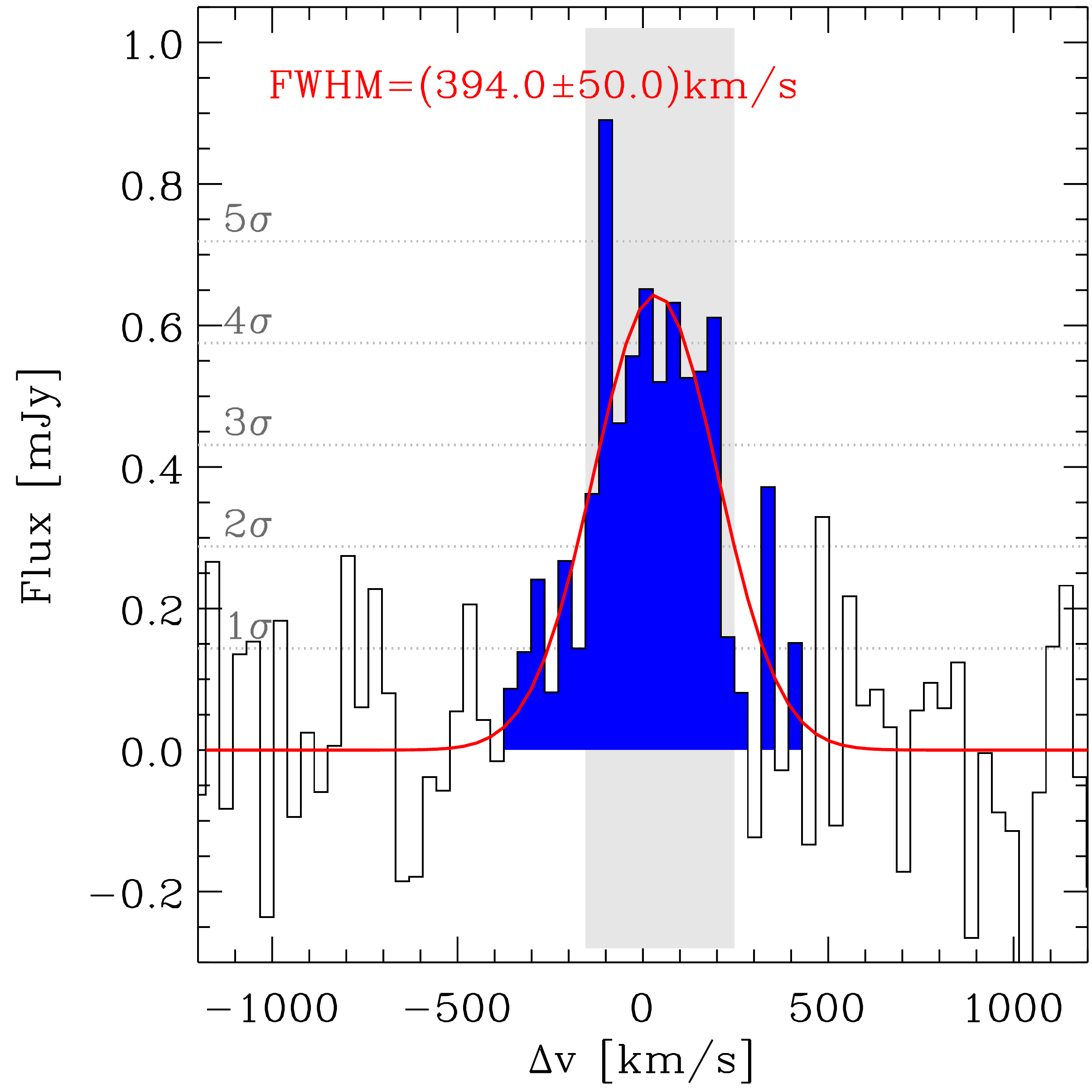}
\end{tabular}
\caption{\emph{Left}: CO(5-4) intensity map, averaged over channels with significant line emission (above 3\,$\sigma$ level, shaded grey area in the right-hand panel). Contours denote the [3, 6, 9]$\times \sigma$ levels. Grey crosses highlight the position of the clumps identified in Fig.~\ref{fig:UDF6462Stamps} and the inset shows the size of the synthesized beam.  Cleaning over the galaxy region was performed employing the Schwab variant of the Clark algorithm and a $0.5\sigma$ cleaning threshold \citep[\textsf{MX} task,][]{Clark_1980,Schwab_1984}. \emph{Middle}: composite $b_{435}-z_{850}-H_{160}$ optical image with ALMA CO(5-4) contours overlaid. \emph{Right}: spectrum extracted over the nuclear region through $uv$-fitting. Velocity offsets are calculated from the redshifted CO(5-4) line emission. Dotted horizontal lines indicate the 1$\sigma$ to 5$\sigma$ noise levels. 
The red line shows the result of a Gaussian profile fit. The FWHM of the CO(5-4) line as derived from the fit is provided in the legend.  We highlight in blue those channels with velocities that are within 2 times the best-fitting Gaussian standard deviation.
\label{fig:COlineInt}}
\end{center}
\end{figure*}

The velocity-integrated CO(5-4) intensity map of UDF6462 is shown in Fig.~\ref{fig:COlineInt}.
 The line emission is concentrated on the nuclear region and detected at a 10\,$\sigma$ significance level (ratio of the peak line flux to the noise in the intensity map).
 
We extracted the total CO(5-4) flux associated with UDF6462 through $uv$-plane fitting with an elliptical Gaussian model, using the \textsc{gildas} task \textsf{UV\_FIT}. 
As initial guesses for the model centre and size, we assumed the coordinates of the peak in the optical images and the ALMA synthesized beam, respectively, and let the parameters vary freely.
The fitting was performed over a velocity range $\Delta$\textit{v}\,$\simeq$\,1680\,km\,s$^{-1}$ (corresponding to a conservative velocity interval equal to $\sim$4.5 times the observed line width derived below).
We obtained a velocity-integrated line intensity of $I_{\rm CO}$\,=\,$S_{\rm CO(5-4)}\Delta$\textit{v}\,=\,$(308.1\pm54.6)$\,mJy\,km\,s$^{-1}$, corresponding to a line luminosity of:
\begin{eqnarray}
L^{\prime}_{\rm CO(5-4)} & = &3.25\times 10^7 S_{\rm CO(5-4)}\Delta \textit{v} \nu^{-2}_{\rm obs} D^2_L (1+z)^{-3} \nonumber \\
 & = & (1.65\pm0.39)\times 10^9{\rm \,K\,km\,s^{-1}\,pc^{-2}} 
\end{eqnarray}

\noindent
where $\nu_{\rm obs}$\,=\,224.23\,GHz, $z=1.57$ and $D_L$ is the luminosity distance in Mpc  \citep{Solomon_VandenBout2005}.
The estimates above have been corrected for the continuum contribution discussed in Section~\ref{sec:Continuum}.
From the best-fitting parameters, we also find a beam-deconvolved FWHM of the CO(5-4) emission equal to $(0\farcs54\pm0\farcs12)\,\times\,(0\farcs20\pm0\farcs13)$, i.e., roughly 2\,kpc in radius. This is in good agreement with the extent of the nuclear region in the optical images.  

\renewcommand{\tabcolsep}{1pt}
\renewcommand{\tabcolsep}{6pt}

We derived the CO(5-4) line spectrum of the nuclear region, presented in the left most panel of Fig.~\ref{fig:COlineInt}, by fitting the $uv$-plane with a Gaussian having parameters fixed to those derived above for the integrated emission. Fitting a Gaussian profile to this spectrum, we found a line velocity width of $\Delta$\textit{v}$_{\rm FWHM}$\,=\,(394$\pm$50)\,km/s. 

\begin{figure}
\begin{center}
\begin{tabular}{c}
\includegraphics[width=0.47\textwidth]{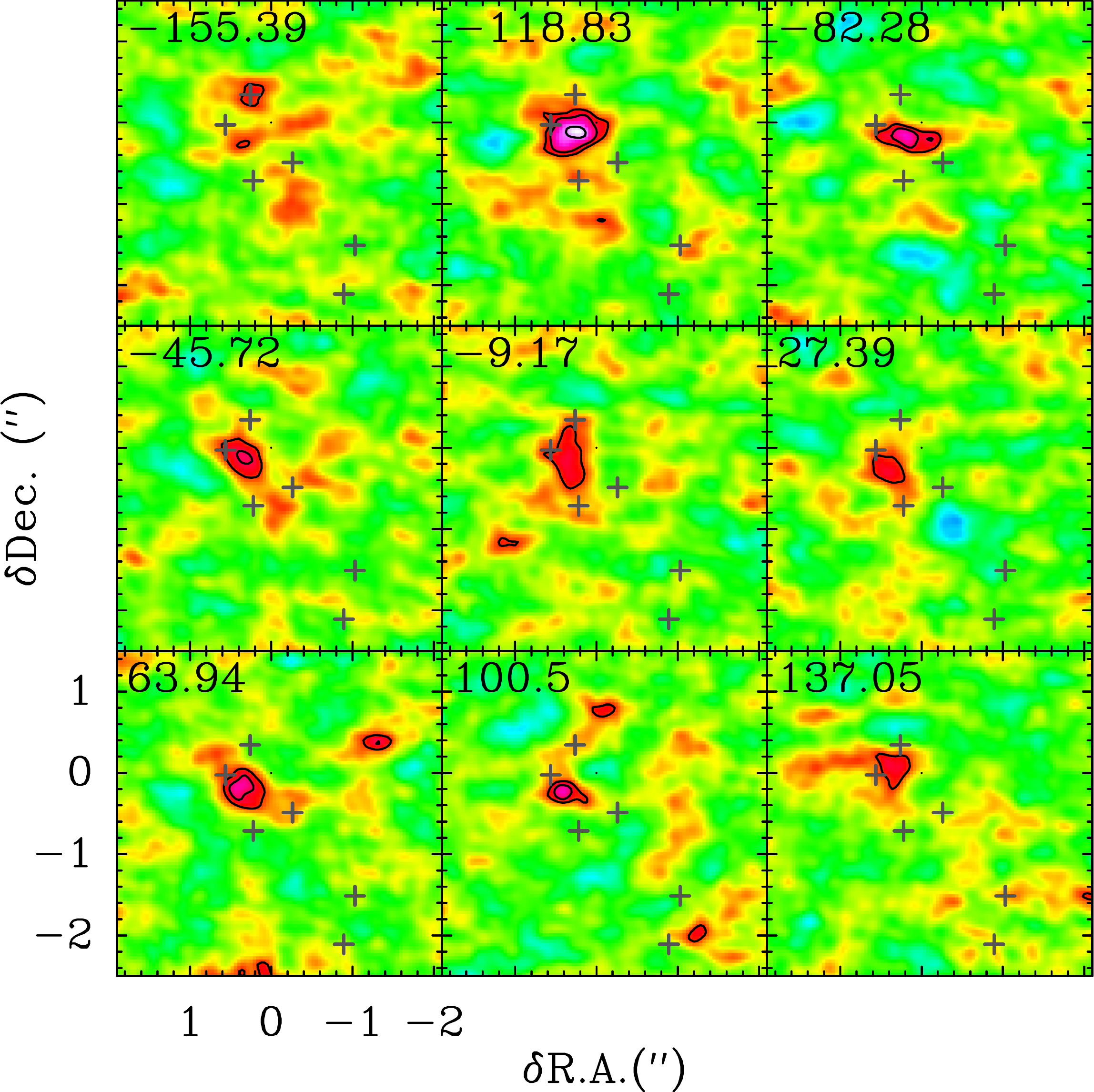} \\
\includegraphics[width=0.47\textwidth]{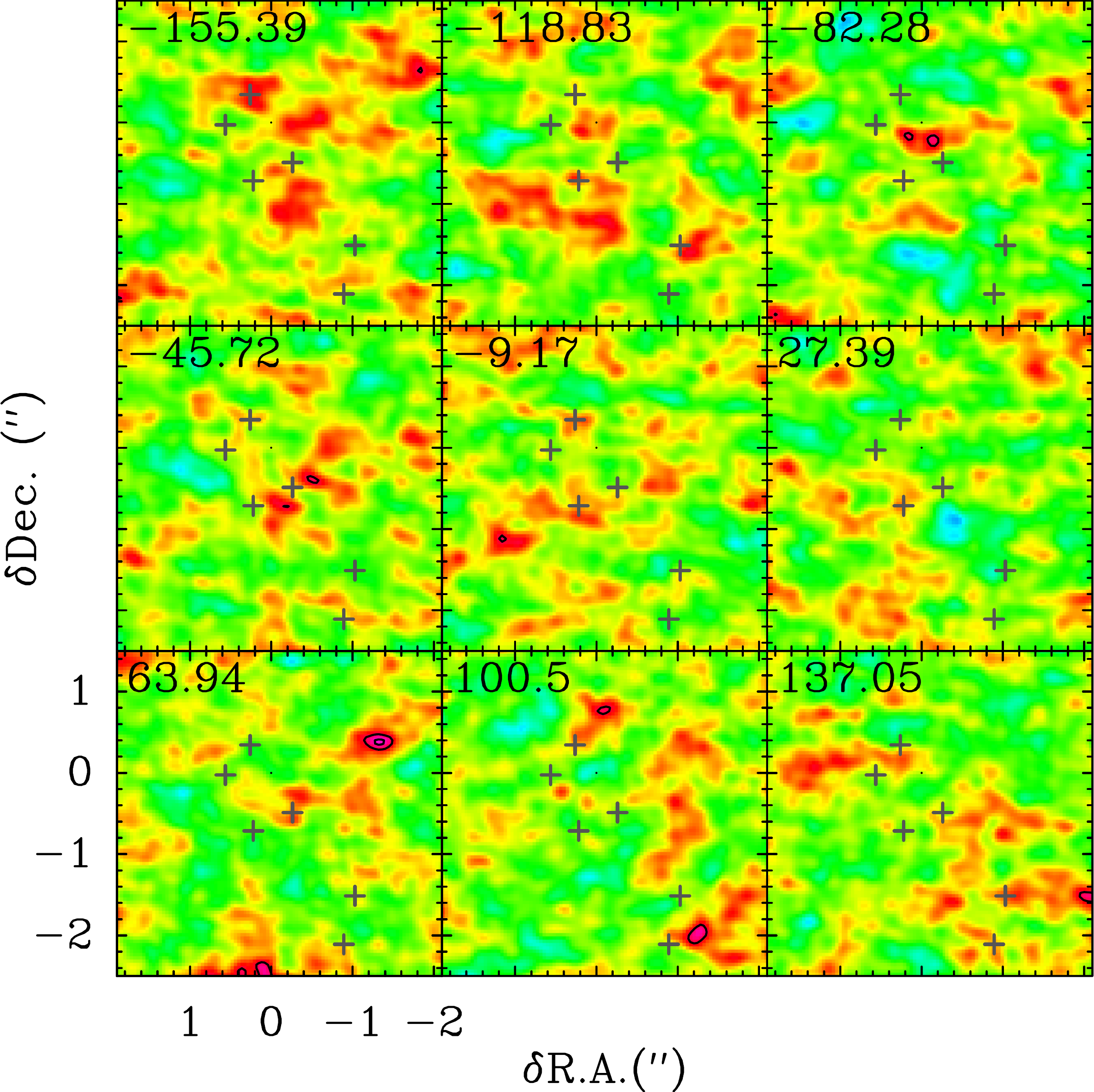} 
\end{tabular}
\caption{\emph{Top:} Cleaned CO(5-4) intensity maps in narrow spectral channels of width 37km/s. Central bin velocities with respect to the galaxy systemic velocity are shown on the top of each panel. Symbols are as in Fig.~\ref{fig:COlineInt}. Contours denote the [3,4,6]$\times \sigma$ levels.
\emph{Bottom:} as above, but after subtraction of the best-fitting Gaussian model for the nuclear region.}\label{fig:UDF6462StampsCube}
\end{center}
\end{figure}

\subsubsection{Clumps}

\begin{figure}
\begin{center}
\includegraphics[width=0.48\textwidth]{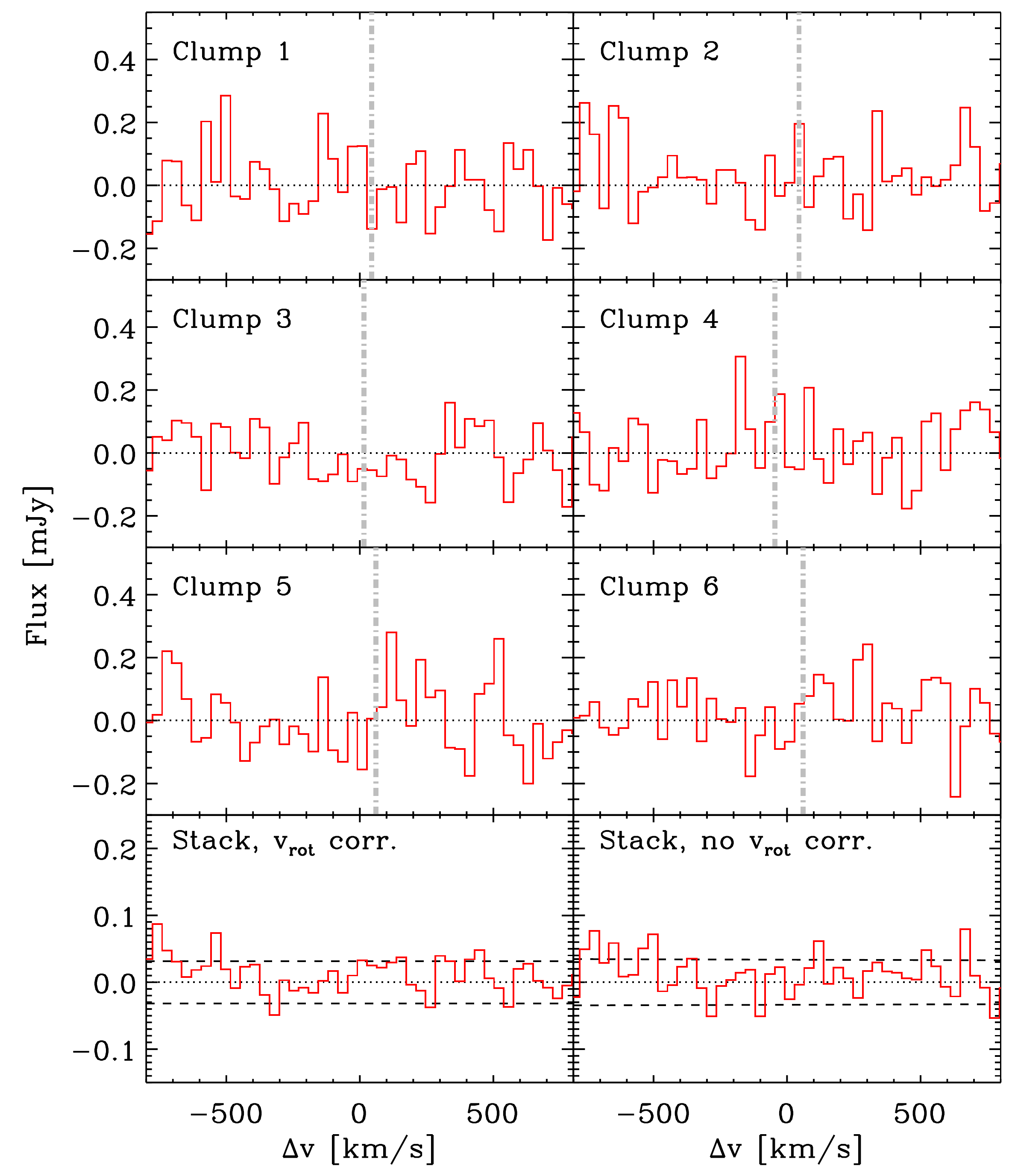} 
\caption{Spectra of the individual clumps in UDF6462.  These spectra are obtained by simultaneously fitting the $uv$-plane with multiple point sources at the clump positions. For this purpose, we employ the residual $uv$ table in which the emission associated with the nucleus has been removed.
To guide the eye, we show with a dotted line the zero intensity level.
 The bottom two rows show the spectra obtained by median stacking the individual clump after applying to each spectrum a velocity shift determined from the $H\alpha$ rotational velocity field (left) or without any shift (right). 
The rotational velocity at each clump position, i.e., the applied shift, is shown with the dash-dotted grey line in the upper six panels.
The horizontal dashed lines indicate the $\pm$1$\sigma$ noise level.}
\label{fig:specClumps}
\end{center}
\end{figure}

None of the the six clumps in UDF6462 is instead detected at the depth of our CO(5-4) map.
Individual ALMA channels at a spectral resolution of 37km/s are shown in Fig.~\ref{fig:UDF6462StampsCube}. 
Even at this high resolution, most of the detected flux is concentrated over the nuclear region with no substantial emission clearly associated with the other six clumps. 
To further verify this, we extracted individual spectra over the clump regions by fitting the visibilities with multiple Gaussian sources having centres and FWHM fixed to those derived from the $z_{850}$ image in Section~\ref{sec:ClumpProp} (similar results are obtained when modelling the clumps as point sources). 
In doing this, it was necessary to remove the contribution of the central component.
For this reason, the extraction was performed on the residual $uv$-table obtained after subtracting the best-fitting elliptical Gaussian model for the nuclear region described above. 
The obtained spectra are presented in Fig.~\ref{fig:specClumps} and confirm the non-detection of the individual clumps.

In order to improve the sensitivity and derive the best possible constraints on the CO content of the clumps, we also produced a stack of the six clump spectra\footnote{Consistent results are obtained when performing stacking in the image plane, rather than on the spectra.}. We employed a mean stacking algorithm, weighted by the flux uncertainties, but consistent results are obtained when performing median stacking. In the work of \citet{Bournaud+08}, it was shown that UDF6264 is rotationally supported and displays a significant velocity gradient.
For this reason, each clump is likely to be displaced with respect to the galaxy systemic velocity by an amount that depends on its location. Prior to stacking, we thus shifted each spectrum by a velocity offset determined from the $H\alpha$ velocity field published in \citet{Bournaud+08}.
To gauge how dependent the stacking results are on this assumption, we also generated another version of the  stacked spectrum by simply taking the median of the spectra without applying any velocity shift. 
The two stacked spectra are shown in the bottom panels of Fig.~\ref{fig:specClumps}; in neither do we detect CO(5-4) line emission clearly associated with the clumps.  

We can use the results of the stacking analysis to place upper limits on the integrated line intensities for the clumps. To do so, an estimate of the line width for the clumps is necessary.
From the $H\alpha$-based velocity dispersion map of UDF6462 presented in \citet{Bournaud+08}, we estimate velocity dispersions of $\sim$\,50\,km/s at the clump positions, resulting in an FWHM of the order 100\,km/s, which we assume to be the typical clump line width.
We note that this is likely a conservative estimate, as the cold gas component traced by CO is expected to have a lower velocity dispersion than $H\alpha$ which is more affected by stellar feedback and outflows. 
Using this line width and the clump flux uncertainty from the multiple Gaussian fitting performed to extract the clump spectra, we find that the $3\sigma$ upper limits for the velocity-integrated line intensities of individual clumps range between $I_{\rm CO(5-4),\rm ul}$\,=\,19 and 25\,mJy\,km\,s$^{-1}$. 
From the stacked spectrum and assuming an average clump FWHM of 0$\farcs$18, we obtain a tighter constraint on the average clump flux equal to $I_{\rm CO(5-4),\rm ul.  stack}$\,=\,9\,mJy\,km\,s$^{-1}$.

Using the upper limit from the stacking analysis, we estimate that each individual clump on average  contributes less than $\sim$3\% to the total CO(5-4) emission in UDF6462, corresponding to roughly a $\lesssim$18\% contribution from all six clumps. For comparison, we estimated in Section \ref{sec:ClumpProp} that $10-40$\% of the total SFR is in the clumps.

\subsection{Rest-frame 500$\mu$m continuum}\label{sec:Continuum}

\begin{figure}
\begin{center}
\includegraphics[width=0.47\textwidth]{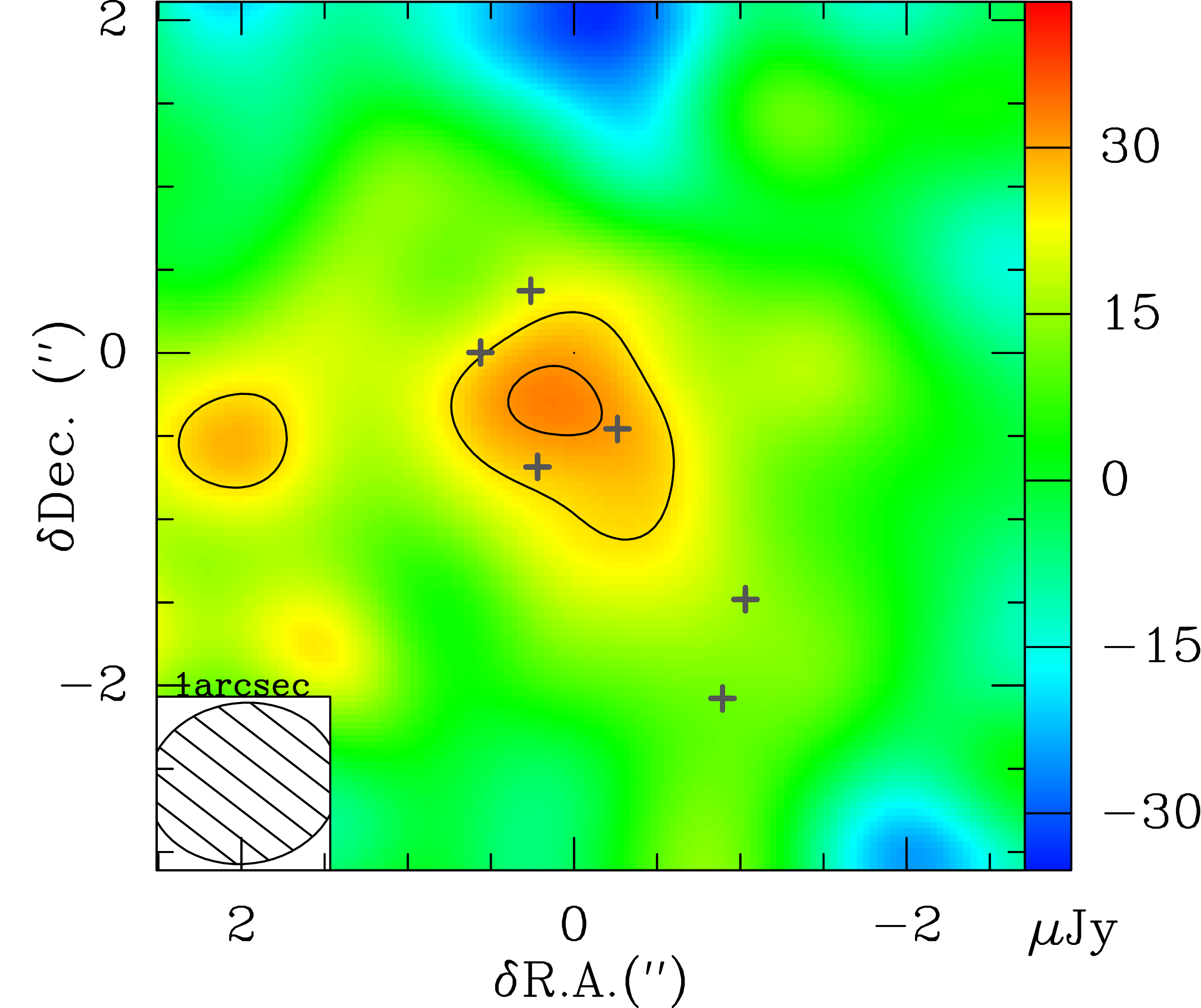}  
\caption{Image of the line-free, 1300\,$\mu$m continuum of UDF6462 tapered to a 1$^{\prime \prime}$ resolution. Contours denote the [2, 2.5]$\times$ noise levels. Grey crosses highlight the position of the clumps identified in Fig.~\ref{fig:UDF6462Stamps}.}
\label{fig:UD6462cont}
\end{center}
\end{figure}

As discussed in Section~\ref{sec:ObsDetails}, together with the line emission, our ALMA observations also cover  7.4 GHz of line-free continuum at a rest-frame wavelength of $\sim$\,500$\,\mu$m.
At the native resolution, mapping of the continuum visibilities is noise-dominated. We therefore generated a  $1^{\prime \prime}$ tapered image that is presented in Fig.~\ref{fig:UD6462cont} and displays diffuse emission centred on the nuclear region of UDF6462 at the $\sim2.5\sigma$ level.
From a $uv$-plane modelling of the (untapered) visibilities with an elliptical Gaussian we obtain a total flux density of $S_{\rm cont}$\,=\,(106.7$\pm$26.2)$\mu$Jy.  (Note the higher SNR in the $uv$-plane, due to the absence of  noise effects introduced by deconvolution algorithms). This estimate of the ALMA continuum flux was used during modelling of the galaxy IR SED in Section~\ref{sec:SFRMassInt} and Fig.~\ref{fig:IRSED}.

From the $uv$-fit we infer an intrinsic size of the continuum emission in UDF6462  of FWHM\,$\simeq$\,($2\farcs$0\,$\pm\,$0$\farcs6)\times(0\farcs$2\,$\pm$\,0$\farcs3)$,  fairly noisy but substantially more extended than the $\sim$\,0$\farcs$5 of the CO(5-4) line.  
To derive an estimate of the continuum associated with the CO-emitting region only, we therefore also performed a Gaussian fit with all parameters, except the total flux, tied to those obtained above for the CO(5-4) line. 
This results in a flux density of (32.1$\pm$10.9)\,$\mu$Jy. The line luminosity quoted in Section~\ref{sec:CO5-4Emission} is corrected for this continuum contribution.\footnote{We directly subtracted this continuum flux from the line integrated flux derived with the Gaussian fit in Section \ref{sec:CO5-4Emission}. However, we have verified that the same result is obtained if subtracting the continuum in the $uv$ space with the \textsc{gildas} \textsf{UV\_SUBTRACT} task and then performing the Gaussian fit on the continuum subtracted $uv$-table. The two methods result in less than 5\% variations in the inferred line luminosity.}

Similarly to what was done in the previous section for the individual clump CO spectra, we also performed a stacking analysis of the continuum image at the clump positions. The clumps remain undetected also in the continuum stack, implying a $3\sigma$ upper limit for the average clump continuum flux density of 12\,$\mu$Jy.


\section{Results}\label{sec:Results} 

We now discuss the implications of the ALMA observations in terms of the molecular gas content and scaling relations for different regions in UDF6462.

\subsection{Star formation, cold dust and CO(5$-$4) distribution}\label{sec:COSFRdist}

It is interesting to compare the distribution of dust and star formation in UDF6462 with that of the dense gas reservoir.
For the SFR, we have independent size estimates from the optical SFR map and the radio data. 
As discussed in Section~\ref{sec:SFRMassInt}, the radio observations indicate an intrinsic FWHM (major axis) size of (0$\farcs$8\,$\pm$\,0$\farcs$ 23).
Although analytical modelling of UDF6462 is made difficult by its complex structure, a \textsc{galfit} S\'ersic or Gaussian fit to the SFR map consistently results in star formation extending over a half-light diameter of  $\sim$\,1$^{\prime \prime}$. The $H\alpha$ intensity map presented in \citet{Bournaud+08} indicate a similar extent of the star formation.
From the 500$\mu$m continuum ALMA image presented in Section~\ref{sec:Continuum} we also infer a significantly extended distribution of the cold dust component over a radius of $\sim$\,(2$\farcs$0\,$\pm$\,0$\farcs$6).
Conversely, from the $uv$-fitting described in the previous section, we found evidence for the CO(5-4) emission to be concentrated over the central 0$\farcs$5.

The question arises whether the size of the CO-emitting region -- especially that of a potentially more extended diffuse component --  could be significantly underestimated as a consequence of noise in the observations. To test this possibility, we simulated ALMA observations of UDF6462 using a sky model generated from the UV-based SFR map, on which we can verify whether observational biases can result in artificially small sizes even when the intrinsic CO(5-4) distribution is as extended as the UV light. While the details of these simulations can be found in Appendix \ref{app:Sim}, we mention here that a model directly created from the UV-based SFR map by applying the SFR-$L_{\rm IR}$ and  $L_{\rm IR}$-$L^{\prime}_{\rm CO(5-4)}$ relations  \citep{Kennicutt_1998,Daddi+15} produces observations with a nucleus that is too faint with respect to the real data by roughly a factor $\times$\,2. 
Two facts contribute to this difference: (1) the scatter in the SFR-$L_{\rm IR}$ and  $L_{\rm IR}$-$L^{\prime}_{\rm CO(5-4)}$ relations, and (2) the fact that the SFR map derived from the \textit{HST} images tends to underestimate dust extinction in the galaxy centre (see Appendix \ref{app:ClumpMeasurements} and discussion in Section~\ref{sec:NuclearRegion}).
To generate simulations that closely reproduce the observed properties of UDF6426, we therefore renormalized the input SFR map to match the observed total line flux over the nuclear region.

We processed the simulated data for the extended CO(5-4) emission exactly as the real ALMA observations (channel averaging, $uv$-plane fitting, etc.), and found that we were able to recover the intrinsic $\sim$\,$1^{\prime \prime}$ size of the input model and its total flux, despite the fact that, just like in the real data, only the central region was clearly visible in the simulated images (see Fig.~\ref{fig:App:FakeObs} in Appendix \ref{app:Sim}). This was also confirmed through tapering of the actual CO(5-4) data which did not reveal emission on larger scales.
We interpret the results of these tests as an indication that the CO(5-4) is truly more centrally concentrated than other star formation tracers or dust in UDF6462 and that we are recovering most of the CO(5-4) emission.

\subsection{$L^{\prime}_{\rm CO(5-4)}$ and $L_{IR}$ correlation} \label{sec:LCO_LIR}

\begin{figure}
\begin{center}
\includegraphics[width=0.48\textwidth]{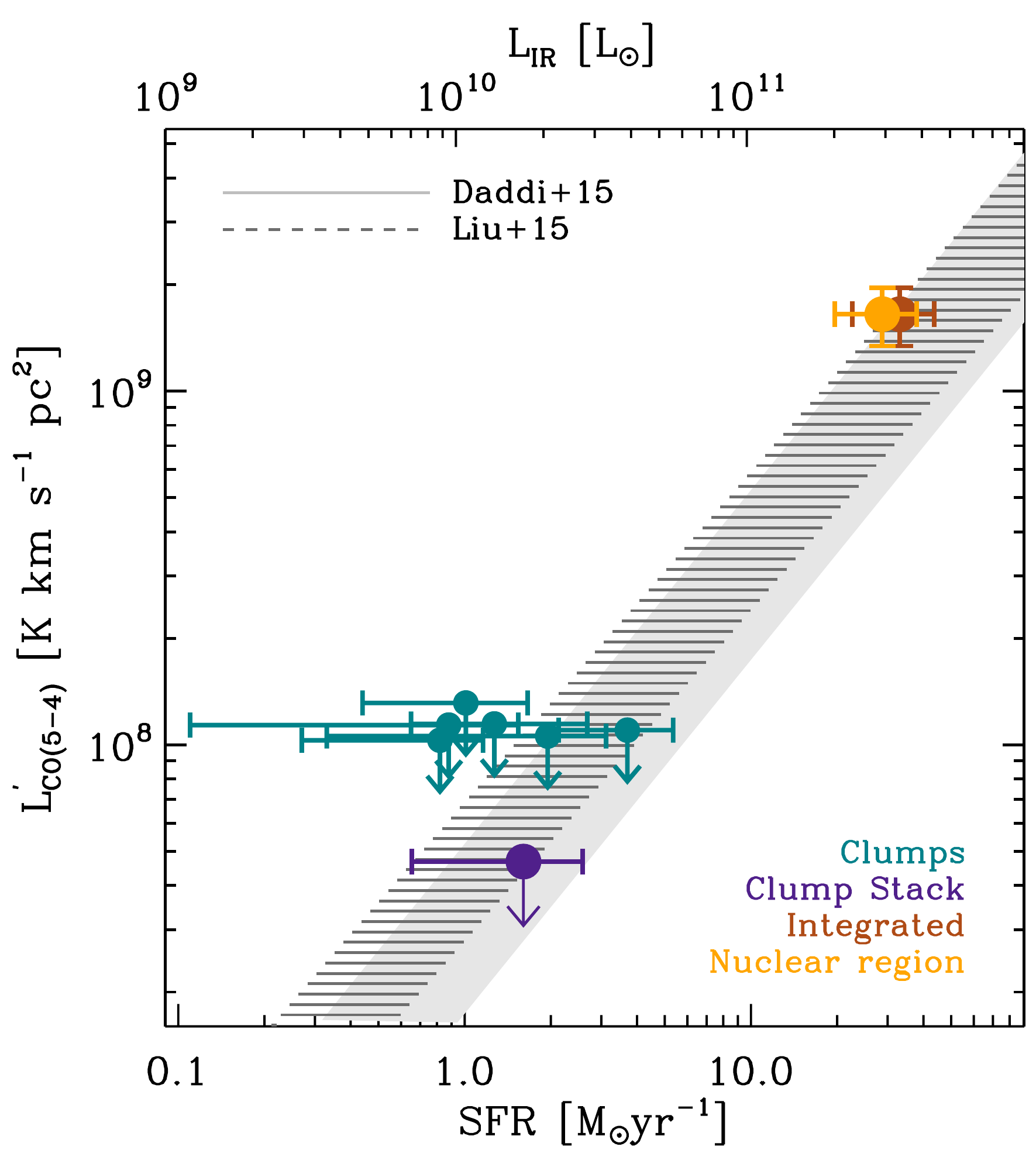}  
\caption{Relation between $L^{\prime}_{\rm CO(5-4)}$ line luminosity and SFR or IR luminosity for resolved regions in UDF6462.
The position of the nuclear region based on the observed  CO(5-4) line flux and SFR from the radio map is shown with the orange symbol.
As we argue in Section~\ref{sec:COSFRdist}, the CO(5-4) detected over the nuclear region likely corresponds to most of the CO(5-4) reservoir in UDF6462. Under this assumption, we can compare the line luminosity with the galaxy integrated $L_{IR}$ luminosity, as illustrated with the brown point.
 Blue symbols are upper limits for the $L^{\prime}_{\rm CO(5-4)}$ in the individual clumps and the purple point is the constraint on the average clump $L^{\prime}_{\rm CO(5-4)}$ that we derive from the stacking analysis. The shaded areas represent the mean and the typical scatter of the CO(5-4)-IR correlation discussed in \citet{Daddi+15} and \citet[][accounting for a factor 1.3 conversion between FIR and total infrared luminosity]{Liu+15}. }
\label{fig:LirLCO}
\end{center}
\end{figure}

As discussed in Section~\ref{sec:intro}, the CO(5-4) transition is believed to be a good tracer of the star-forming gas, resulting in an almost linear CO(5-4)--IR correlation that holds over more than 5 orders of magnitude and for a variety of systems, including local galaxies as well as high redshift sub-mm galaxies.
It is thus interesting to verify where the UDF6462 clumps and nuclear region lie on the $L_{IR}$ and $L^{\prime}_{\rm CO(5-4)}$ plane. This is shown in Fig.~\ref{fig:LirLCO}, together with the best-fitting relations and associated scatter from the works of \citet{Daddi+15} and \citet{Liu+15}.
For regions within the galaxy we do not have a direct measurement of the IR emission, due to the low angular resolution of the \textit{Spitzer}/\textit{Herschel} IR data. Instead, we use the clump and nuclear region SFR (from the SFR map and the VLA 6GHz radio observations, respectively)  as a proxy for their IR flux. 

The CO(5-4) measurement for the nuclear region places it on the correlations found in the existing literature. 
As we argued in the previous section, we believe that the CO(5-4) detected over this region likely corresponds to a large fraction of the total CO(5-4) in this galaxy. 
By assigning to the CO(5-4) luminosity (measured over the nuclear region) the galaxy total IR luminosity from the IR-SED fit in Section~\ref{sec:SFRMassInt}, we can verify where the entire galaxy would lie if this assumption is true. This is illustrated with the brown symbol, showing that the integrated measurement would be fully consistent with the typical  CO(5-4)--IR relation.
 At the depth of our observations, we can only place upper limits on the molecular gas content in the clumps, either individually or through the stacking analysis. 
These limits are consistent with clumps also lying within the current measurements of the  CO(5-4)--IR correlation.

\subsection{Gas mass and star formation efficiency}\label{sec:Mgas_SFE}

From the ALMA data, we can derive estimates of the gas content in UDF6462 using different approaches.
We note that most of the techniques described below effectively yield a total gas mass ($M_{\rm gas}$\,=\,$M_{\rm mol}$+$M_{\rm HI}$), with the exception of gas masses inferred from the CO(5-4) line emission that correspond to molecular gas only ($M_{\rm mol}$\,=\,$M_{\rm H_{2}}$+$M_{\rm He}$). 
From now on we will make the assumption that the molecular phase dominates the total gas content, i.e., $M_{\rm gas}$\,$\simeq$\,$M_{\rm mol}$ as found observationally and theoretically for high redshift galaxies \citep{Obreschkow_Rawlings2009,Daddi+10a,Tacconi+10}, and consequently that the different estimates can be directly compared without applying any correction for atomic gas. 

\subsubsection{Gas mass from sub-mm continuum}

The ALMA continuum observations described in Section~\ref{sec:Continuum} probe the Rayleigh--Jeans tail of the dust emission, enabling us to constrain the shape of the IR SED and perform the IR multicomponent fit outlined in Section~\ref{sec:SFRMassInt} (see \citealt{Magdis+2011,Magdis+12} for details on the method). 
From this fit a dust mass, $M_{\rm dust}$, is calculated and can be converted into a total (H$_{2}$+H{\small I}) gas mass, $M_{\rm gas}$, through a gas-to-dust ratio, $M_{\rm gas}$\,=\,$\delta_{\rm GDR} M_{\rm dust}$. We inferred the latter using the $\delta_{\rm GDR}$--metallicity relation in \citet{Magdis+2011} and assuming the average galaxy metallicity to be $12+\log(O/H)$\,=\,8.53, as measured in \citet{Bournaud+08} and expressed in the \citealt{Pettini_Pagel04} system. We find a ratio $\delta_{\rm GDR}$\,$\simeq$\,120 for UDF6462.

The results of these calculations are summarized in Fig.~\ref{fig:SK} where we show the position of UDF6462 and regions therein on the integrated Schmidt-Kennicutt plane \citep{Schmidt_1959,Kennicutt_1998b}.
From the gas-to-dust/SED fitting technique we calculated for the entire galaxy a dust mass of log($M_{\rm dust}/\Msol$)\,=\,8.22$\pm$0.41 and a $M_{\rm gas, gal}$\,=\,$2.0^{+4.4}_{-1.4}\times 10^{10}\Msol$, corresponding to a gas fraction  $f_{\rm gas, gal}$\,=\,$M_{\rm gas}/(M_{\rm gas}+M_{\star})$\,=\,0.45 and  a depletion time-scale $\tau_{\rm depl, gal}$\,=\,0.51\,Gyr (equivalent to a star formation efficiency SFE\,=\,SFR/M$_{\rm gas}$\,=\,1.96\,Gyr$^{-1}$).  This depletion time-scale is in good agreement with those typically found in MS galaxies with similar masses and redshifts as UDF6462 \citep[e.g.][]{Daddi+10b,Tacconi+10}.
The MS-like behaviour is also reflected by an average radiation intensity of $\langle U \rangle$\,=\,16, as derived from the IR-SED fitting.

Due to the large point spread function (PSF) of the IR observations, the full SED approach can only be applied to the entire galaxy, with no possibility of distinguishing the individual clumps.
However, empirical relations linking monochromatic continuum fluxes to total gas masses are available in the literature \citep[e.g.][]{Scoville+14,Groves+15} which we can apply to place an upper limit on the clump gas mass using the high-resolution, 500\,$\mu$m ALMA data alone. In particular, using the relation between 500\,$\mu$m luminosity and gas mass derived in \citet{Groves+15} and the clump continuum flux upper limit derived in Section~\ref{sec:Continuum}, we obtain a 3$\sigma$ upper limit for the gas mass in the average clump of $M_{\rm gas, cl}$\,=\,$3.5 \times 10^{9}\Msol$ (light purple symbol in Figure  \ref{fig:SK}). 
For comparison, applying the same monochromatic technique to the integrated galaxy we calculate a total gas mass of $(3.0\pm0.8)\times 10^{10}\Msol$.

\subsubsection{Molecular gas mass from CO(5-4) emission}\label{sec:GasFromCO}

\begin{figure*}
\begin{center}
\begin{tabular}{cc}
\includegraphics[width=0.48\textwidth]{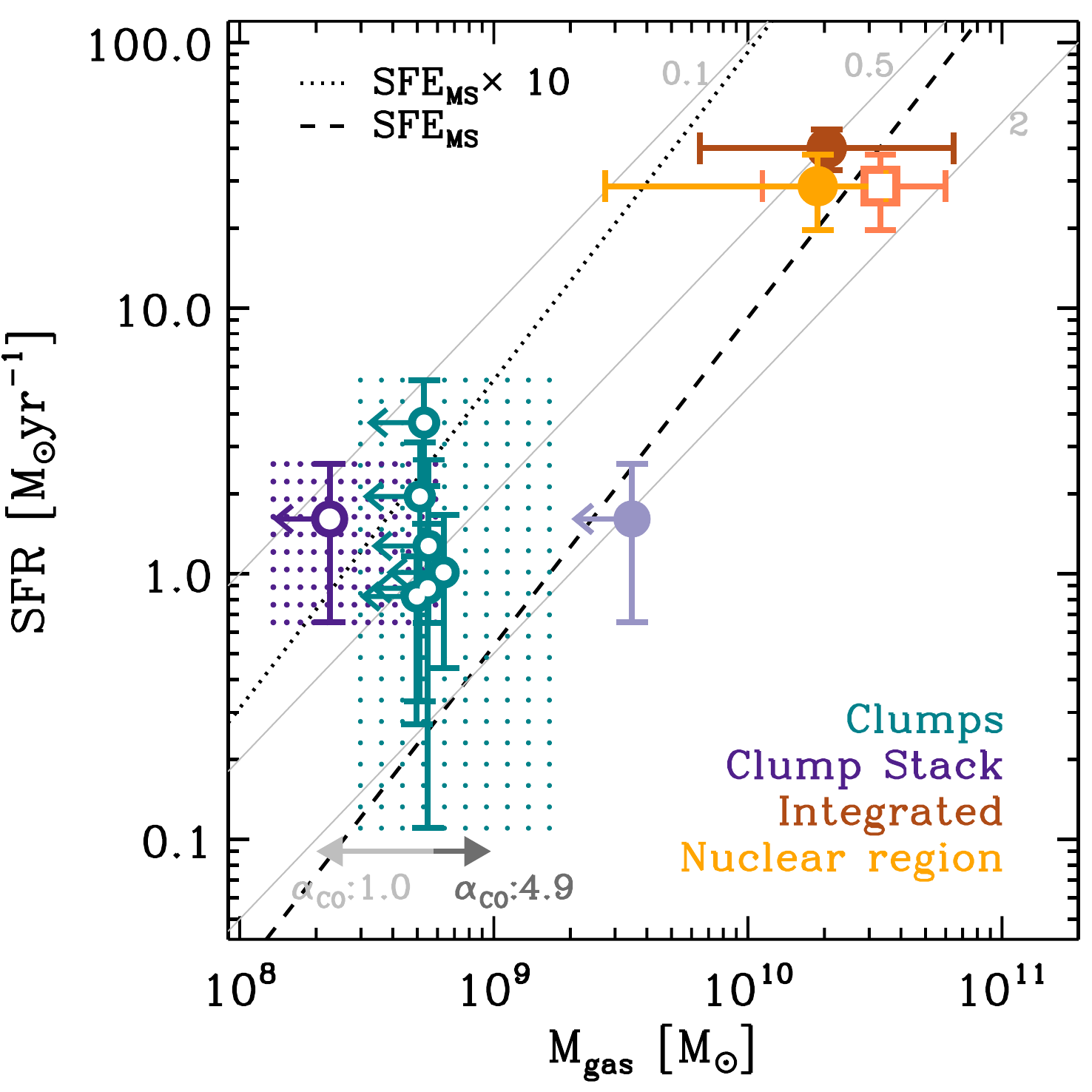}  & \includegraphics[width=0.48\textwidth]{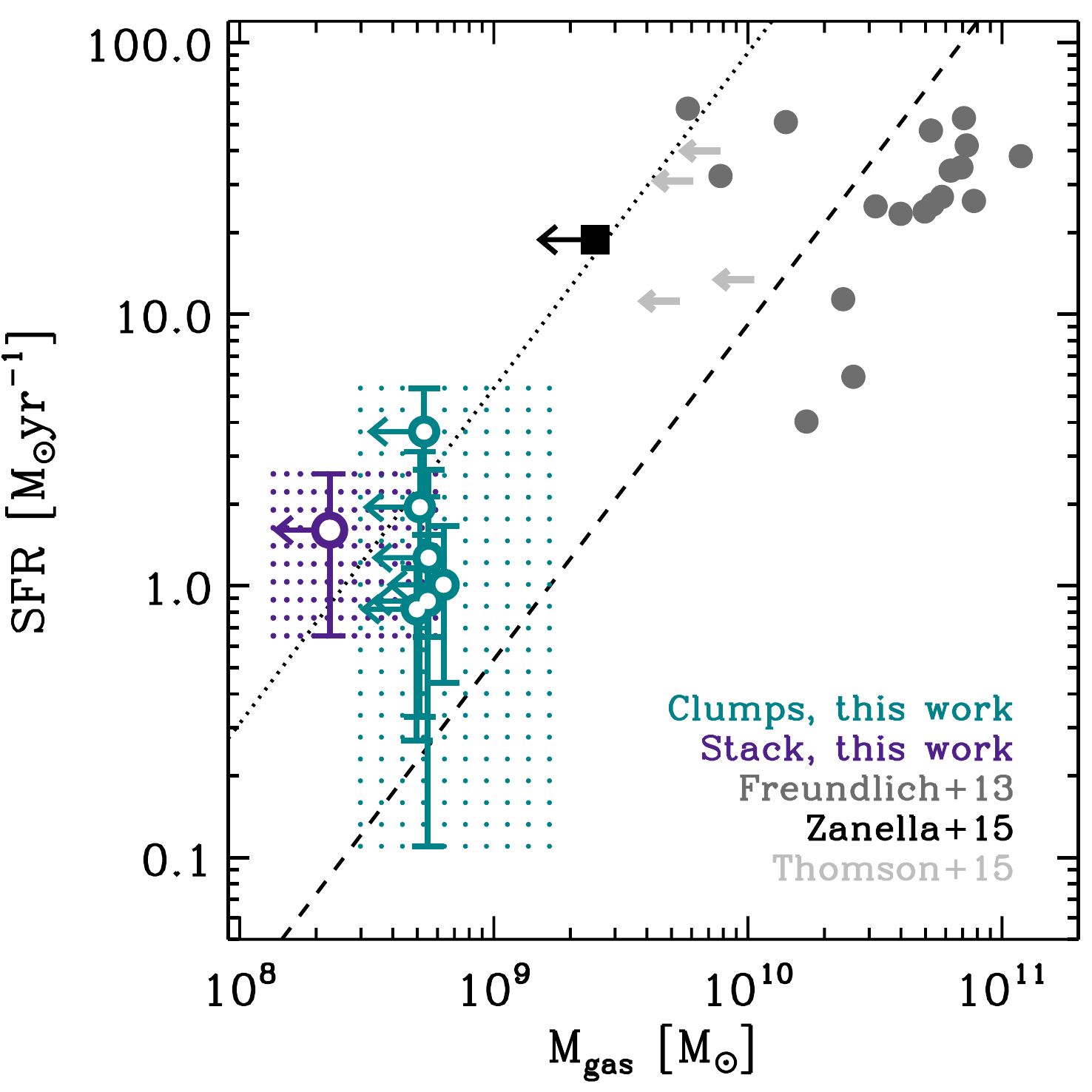} \\
\end{tabular}
\caption{Position of UDF6462 and regions within it on the integrated Schmidt-Kennicutt plane. 
\emph{Left:} The brown symbol indicates the value of $M_{\rm gas, gal}$ for the entire galaxy derived from the IR SED fitting with the \citep{Draine_Li_2007} models. The filled orange circle is the molecular gas mass for the nuclear region calculated from the ALMA CO(5-4) data ($M_{\rm mol, nuc}$). The nuclear region gas mass inferred from the dynamical analysis is shown as the empty orange square ($M_{\rm gas, nuc}$). Turquoise symbols with arrows represent upper limits on the molecular gas masses of individual clumps calculated assuming $\alpha_{\rm CO}$\,=\,2.9$\Msol/({\rm \,K\,km\,s^{-1}\,pc^{-2}})$ and $R_{51}$\,=\,0.6 \citep{Bournaud+15}.
The dark purple point is the average clump gas mass inferred from stacking of the CO(5-4) observations ($M_{\rm mol, cl}$).
The dotted areas illustrate how these upper limits would change if $R_{51}$\,=\,1 (thermally excited CO(5-4); left-hand edge of dotted region) or $R_{51}$\,=\,0.23 (observed for high-$z$ MS galaxies, \citealt{Daddi+15}; right-hand edge).
The effect of a different $\alpha_{\rm CO}$ is shown with the grey arrows in the lower left corner. The light purple point is the upper limit on the total gas mass obtained from continuum stacking ($M_{\rm gas, cl}$). The dashed line highlights the locus of MS galaxies as parametrized in \citet{Sargent+14} and the dotted line a factor $\times 10$ enhancement in SFE. Thin grey lines with numbers are lines of constant depletion time (in Gyr).
\emph{Right:} comparison of clump properties in UDF6462 with some recent literature measurements for resolved regions within galaxies: the young clump in \citet{Zanella+15}; clump ensembles over a $\sim$8\,kpc region from \citet{Freundlich+13} and CO-identified clumps in the sub-mm galaxy of \citet{Thomson+15} (adapted from their fig. 7 and table 2).}
\label{fig:SK}
\end{center}
\end{figure*}

We can also use the CO(5-4) data to calculate molecular gas masses, $M_{\rm mol}$, for the nuclear region and the individual clumps in UDF6462. 
We stress here that the derivation of the total molecular gas content, to which a more diffuse component than the dense gas traced by CO(5-4) could also contribute, involves two steps that are subject to large uncertainties:  the conversion of $L^{\prime}_{\rm CO(5-4)}$ into a CO(1-0) luminosity (for which an excitation correction $R_{51}$ must be adopted) and the conversion of CO(1-0) luminosity into molecular mass (for which a specific mass-to-light ratio -- the CO-to-$H_2$ conversion factor, $\alpha_{\rm CO}$ -- must be chosen).
To highlight the systematics involved in this estimate, we carry out the calculation of $M_{\rm mol}$  for a range of plausible values for $R_{51}$ and $\alpha_{\rm CO}$ and show in Fig.~\ref{fig:SK} the corresponding variation in $M_{\rm mol}$.

For the nuclear region -- which, as illustrated in Fig.~\ref{fig:MSClumps}, has a slightly enhanced sSFR compared to integrated measurements for MS galaxies -- we calculated $M_{\rm mol}$ employing either MS- or SB-like excitation correction and $\alpha_{\rm CO}$ value. In the first case, we used an excitation correction $R_{51}=0.23$ as derived in \citet{Daddi+15} for normal, MS galaxies at a similar redshift as UDF6462, together with a roughly Milky Way $\alpha_{\rm CO}$\,=\,4.87$\Msol/({\rm \,K\,km\,s^{-1}\,pc^{-2}})$.
We calculated this conversion factor by applying the relation between metallicity and $\alpha_{\rm CO}$ in \citet{Wolfire+10} and using the galaxy average metallicity $12+\log(O/H)$\,=\,8.53\footnote{This corresponds roughly to a solar metallicity, if we correct for the offset between the \citet{Pettini_Pagel04} metallicity system and other metallicity systems in which Milky Way mass galaxies have a metallicity in good agreement with the directly measured solar metallicity value $12+\log{\rm (O/H)}$\,=\,8.66  \citep{Asplund+05}.}. 
For the SB case, we instead used $\alpha_{\rm CO}$\,=\,1 and an excitation correction equal to $R_{51}$\,=\,0.6 as observed in local/high-$z$ SB \citep[e.g.][]{Downes_Solomon1998,Magdis+12,Greve+14}.
The average of these two estimates yield a molecular gas mass equal to $M_{\rm mol, nuc}$\,=\,$(1.9\pm1.6)\times 10^{10}\Msol$ (shown as a filled orange circle in Fig.~\ref{fig:SK}; the error bar indicates the spread in these estimates), corresponding to a gas fraction $f_{\rm gas, nuc}$\,$\simeq$\,0.7 in the nuclear region.

As a consequence of the lack of observations of the ISM at the clump scale, it is unknown which conversion and excitation correction factors should be applied to the clumps. Consequently, the precise location of the clumps in the Schmidt-Kennicutt plane is uncertain. 
There are, however, predictions from numerical simulations that we can use to better pinpoint their properties.
In the simulations of \citet{Bournaud+15} giant star-forming clumps in high-redshift galaxies are responsible for high-$J$ excitation and have CO SLEDs that are more excited than their host galaxy, with a clump excitation corrections of $R_{51}$\,=\,0.6. 
For comparison, $R_{51}$ values typically found in local (Ultra) Luminous Infrared Galaxies (ULIRGs) range between 0.4 and 0.8 (see, e.g., extrapolation from lower-$J$ measurements provided in \citealt{Papadopoulos+12} and the average compilation of this data in \citealt{Daddi+15} giving $R_{51}$\,$\simeq$\,0.5). A similar range is observed for high redshift, bright sub-mm galaxies (e.g., $R_{51}$\,=\,0.32 in \citealt{Bothwell+13} and $R_{51}$\,=\,0.83 in \citealt{Riechers+11}), whereas \citet{Daddi+15} report a  $R_{51}$\,=\,0.23 in normal, MS galaxies at z\,$\simeq$1.5 and a $R_{51}$\,=\,0.08 is measured for the Milky Way \citep{Fixsen+99,Carilli_Walter2013}.
Moreover, in these simulations giant clumps exhibit CO-to-H$_2$ conversion factors that are in between those of merger-induced SBs and normal galaxies we quoted above. In particular, from fig. 5 of \citet{Bournaud+15} we derive an $\alpha_{\rm CO}$\,=\,2.9$\Msol/({\rm \,K\,km\,s^{-1}\,pc^{-2}})$ for the clumps.

The  upper limits on clump gas masses arising from these simulation-motivated choice of parameter ($R_{51}$\,=\,0.6 and $\alpha_{\rm CO}$\,=\,2.9$\Msol/({\rm \,K\,km\,s^{-1}\,pc^{-2}})$) are shown as turquoise symbols in Fig.~\ref{fig:SK} and the limit derived from the stacked spectrum is plotted as the dark purple point.
To highlight the uncertainties on the parameter choice, we illustrate with the dotted areas how the estimated limits on the clump gas masses vary for other possible values of $R_{51}$. Similarly, the effect of variations in the $\alpha_{\rm CO}$ is shown with the grey arrows. 
As shown in the right-hand panel of this figure, our measurements probe roughly a factor of 10 lower SFR and gas masses with respect to other observations of resolved regions in z\,$>$\,1 galaxies.

Applying the fiducial values of $R_{51}$ and $\alpha_{\rm CO}$ from \citet{Bournaud+15}, we find that clumps in UDF6462 have molecular masses of $\simeq$\,$5 \times 10^{8}\Msol$ and depletion time-scales $\simeq$\,0.1-0.6\,Gyr (SFE in the range $\simeq$\,1.6-6\,Gyr$^{-1}$).
Although some clumps are consistent, within uncertainties, with the SFR-$M_{\rm gas}$ relation of MS galaxies, there is a general tendency for them to be offset from it to higher SFE. This is especially clear for the stacking analysis, which provides a limit on average clump depletion time-scale in the range $\tau_{\rm depl, cl}$\,$\simeq$\,0.15\,Gyr (0.1-0.4\,Gyr for the different choices of $R_{51}$ and $\alpha_{\rm CO}$, see dark dotted area). From the stacking we also derive an average clump gas mass of $M_{\rm mol, cl}$\,$\simeq$\,$2\times 10^{8}\Msol$ (ranging between $1-6\times 10^{8}\Msol$) and gas fraction of $f_{\rm gas, cl}$\,$\simeq$\,30\% (20-60\%).
The upper limit on the average $M_{\rm gas, cl}$ of the clumps from the continuum stacking is consistent with this estimate, but is too shallow to improve the constraints from the CO(5-4) measurement.

We note that a short depletion timescale/high efficiency in clumps is not in contradiction with the fact that  they do not display a significant offset from the $L^{\prime}_{\rm CO(5-4)}$--$L_{IR}$ followed by the nuclear region and integrated galaxies (see Section~\ref{sec:LCO_LIR}). As discussed in \citet{Daddi+15}, while normal galaxies and high-efficiency SBs display different $L^{\prime}_{\rm CO}$/$L_{IR}$ ratios when low-$J$ transitions are considered, they have almost indistinguishable $L^{\prime}_{\rm CO(5-4)}$/$L_{IR}$ ratios. This is interpreted as an indication that the CO(5-4) transition almost directly traces the  gas phase participating in star formation, independently of whether the latter proceeds in high a efficiency mode or not.

\subsubsection{Dynamically-inferred gas mass}\label{sec:GasMassDyn}

Finally, for the nuclear region we can also infer the gas mass using dynamical arguments. An approximate estimate of the dynamical mass in this region can be calculated using either the isotropic virial estimator \citep{Spitzer1987}:
\begin{equation}
M_{\rm dyn, vir}(r<r_{\rm CO})=\frac{5\sigma_v^2 r_{\rm CO}}{G} = 2.0\times 10^5\, \Delta \textit{v}^2_{\rm FWHM}r_{\rm CO},
\end{equation}

\noindent
or the relation for high-$z$, clumpy disc galaxies in \citet{Daddi+10a} which was calibrated on simulations:

\begin{equation}
M_{\rm dyn, high-z}(r<r_{\rm CO})=1.3\times  \frac{(\Delta \textit{v}_{\rm FWHM}/2)^2 \times r_{\rm CO}}{G \sin^2i}\,.
\end{equation}
\noindent
Here the numerical factor accounts for biases affecting the dynamical mass estimate in these clump-dominated systems\footnote{This factor was derived within the effective radius by these authors, but we assume here that the same correction can be applied to any radius.}.

In the equations above, $r_{\rm CO}$\,=\,0$\farcs$27\,$\simeq$2\,kpc is the half width at half-maximum of the best-fitting Gaussian for the CO(5-4) emission, $\Delta$\textit{v}$_{\rm FWHM}$\,=\,2.35$\times \sigma_v$\,=\,394\,km/s is the line velocity width from the Gaussian fit to the integrated spectrum and we assumed an inclination of $i$\,=\,63 as derived from the dynamical modelling of \citet{Bournaud+08}. We use the average of these two estimates 
$M_{\rm dyn} (r<r_{\rm CO})$\,=\,(5.36$\pm$2.0)$\times10^{10}$\,$\Msol$ as our fiducial value (the error represents the range of values spanned by the two approaches). 
This is in good agreement with the dynamical mass of $\sim 5 \times10^{10}$\,$\Msol$ inferred in \citet{Bournaud+08} within the optical radius.  We then calculate an approximate total gas mass in the nuclear region following an approach similar to that in \citep{Daddi+10a,Tan+14}:
\begin{equation}
M_{\rm dyn}(r<r_{\rm CO})=f_{\star}M_{\star}+M_{\rm gas}(r<r_{\rm CO})+M_{\rm dark}(r<r_{\rm CO})
\end{equation}

\noindent
where $f_{\star}$ is the the fraction of the total galaxy stellar mass $M_{\star}$ within $r_{\rm CO}$ and $M_{\rm dark}$ is the dark matter component within the same radius.
We calculated $f_{\star}$ from our resolved mass maps and found it to be $f_{\star}\simeq 0.5$. The dark matter contribution is uncertain and we hence derive  $M_{\rm gas}$ as the average of two extreme assumptions: $M_{\rm dark}$\,=\,0, for a baryon-dominated nucleus as suggested by recent observations at $z\gtrsim 1$  \citep{Wuyts+16}, and $M_{\rm dark}$\,=\,0.25$M_{\rm dyn}$ typically adopted within the half-light radius of high redshift galaxies \citep{Daddi+10a}.
 
From the dynamical considerations above, we infer a gas mass in the nuclear region of $M_{\rm gas, nuc}$\,=\,$3.3^{+2.7}_{-2.2} \times 10^{10}\Msol$.
Taken at face value, the dynamical and CO(5-4) gas masses support a picture in which the nuclear region is gas dominated ($f_{\rm gas, nuc}$\,$>$0.5).
From the average of the two estimates, we infer $\tau_{\rm depl, nuc}$\,$\simeq$\,1\,Gyr.


\section{Summary and Discussion}\label{sec:Summary}

In this paper we have presented ALMA observations of the CO(5-4) transition in a clumpy, MS galaxy at $z$\,$\simeq$\,1.5 and located in the HUDF field in GOODS-S.
Our observations have a resolution of $\sim$0.5$^{\prime\prime}$ which enables us to spatially resolve the CO(5-4)-emitting regions within the galaxy and to study the dense, star-forming gas  at a scale that is comparable to the individual giant clumps which are visible in \textsc{HST}, rest-frame UV images.
Together with the CO(5-4) line we also probe the underlying, rest-frame 500\,$\mu$m continuum tracing the cold dust component and providing constraints on the total gas reservoir.

We can summarize our results as follows:
\begin{enumerate}
\item UDF6462 displays integrated gas properties that are fully consistent with its MS-like nature. From the ALMA continuum observations, we derive a $\tau_{\rm depl, gal}\simeq$0.5\,Gyr and a mean radiation field intensity $\langle U \rangle$\,=\,16, which are within the observed range of normal, z\,$>$\,1 galaxies.
As expected given its clumpy appearance and as typically observed at this redshift, UDF6462 has a high gas fraction of $f_{\rm gas, gal}$\,$\simeq$\,0.5.

\item The bulk of the CO(5-4) emission is cospatial with the nuclear region of the galaxy,
where dust obscured star formation is taking place (possibly building the bulge component of the galaxy) and where a central mass concentrations is present.
Calculations based on both the CO(5-4) emission and dynamical considerations suggest a high gas fraction of $f_{\rm gas, nuc}$\,$\simeq$\,0.7 and a  $\tau_{\rm depl, nut}\sim$1\,Gyr for this nuclear region.

\item The distribution of CO(5-4) line emission in UDF6462 appears less extended than star formation, as traced by the UV or radio images, or the distribution of cold dust, as traced by the ALMA continuum observations.
From the analysis of simulated ALMA observations, we conclude that this difference is intrinsic and not caused by observational biases.
 
\item CO(5-4) emission from individual clumps within UDF6462  is not detected but we can place upper limits on their CO(5-4) luminosity, either individually or on average, through a stacking analysis.
These limits are consistent with clumps falling on the common relation between $L^{\prime}_{\rm CO(5-4)}$ and $L_{IR}$ (or SFR) that holds locally and at high redshift for normal and SB galaxies. 

\item With the caveat that such a calculation carries a number of assumptions, we also derived limits for the clump molecular gas reservoirs using theoretically-motivated conversion factors.
From the stacking analysis  we infer a relatively low gas fraction of the order of $\simeq$\,30\% at the scale of giant clumps, with a plausible range of 20--50\%. This is indicative of clump gas fractions that are not higher than that of their host galaxy and suggests that clumps may not be gas dominated. 
It also implies that, for their typical SFR, the clumps might have a relatively elevated SFE.
It is interesting to note that, as opposed to SB galaxies, the high SFE  would be associated to a modest, if any, enhancement in sSFR, as indicated by the fact that the clumps fall on the extension of the MS as derived for integrated galaxies (see Figure \ref{fig:MSClumps}).
\end{enumerate}

We now conclude by offering possible physical explanations of our findings.  

\subsection{Scaling relations for giant clumps}

The exact position of the UDF6462 clumps in the Schmidt-Kennicutt plane is uncertain, due to an observationally poorly constrained clump CO-to-$H_2$ conversion factor or excitation correction.
However, with reasonable assumptions for $\alpha_{\rm CO}$ and $R_{51}$ and taking the clump SFR at face value,  molecular gas reservoirs inferred from the CO(5-4) emission 
suggest that clumps in UDF6462 reside in a part of the Schmidt-Kennicutt plane with shorter depletion times/lower gas masses, compared to the integrated measurements of MS galaxies. 

The suggested offset from the MS relation could be interpreted as genuinely high SFE in the clumps (driven by a high SFR) or, instead, reflect an advanced stage of gas depletion.
In the first scenario, clumps would behave as `kiloparsec-scale SB' within their host galaxies. 
Previous indirect evidence for high-efficiency star formation in clumps was found in a young, $<$\,10\,Myr old clump in \citet{Zanella+15}. \citet{Freundlich+13} report a wide range of depletion time-scales for ensembles of clumps ($\sim$\,8\,kpc scale) in four massive, $z$\,=\,1.2 galaxies, with some of the clump ensembles also clearly displaying high efficiencies (see right-hand panel in Fig.~\ref{fig:SK}). Although for other ensembles the SFE measurements can be comparable to those typically found in normal galaxies, we note that, due to the resolution of their observations, the molecular masses (and SFRs) in these ensembles will likely contain contributions from different regions within the diffuse galactic disc, which could lower the measured SFE from that of purely clump-dominated areas.  
Our observations instead enable us to single-out clump regions; the range of clump SFEs observed in UDF6462 could reflect variations in the SFE enhancement during slightly different stages of their evolution, as seen in numerical simulations \citep{Zanella+15}.

In the alternative scenario, a high measured SFE could be related to nearly completed gas consumption, possibly coupled with efficient feedback quickly removing the gas reservoir from the clumps. Strong outflows at the clump scale are seen in observations and simulations \citep[e.g.][]{Genzel+11,Newman+12,Bournaud+14}. Given the characteristic time-scale of $\sim$100-300 Myr for UV emission as an SFR tracer,  the high SFE might be an artefact of a time-scale ``mismatch" with the CO-emission which is an instantaneous tracer of the gas content. In this case the actual SFE towards the end of the depletion process could be lower than we measure.

Finally, we must also consider the possibility that the clump CO(5-4) non-detections and inferred low gas masses simply reflect uncertainties in their SFR, rather than intrinsic gas physics. 
As we discuss in Section~\ref{sec:ClumpProp}, there are systematics that make it difficult to precisely determine the SFR of clumps from broad-band photometry and the SFR estimates for individual clumps can vary by up to $\sim$1 dex.
In a scenario in which the clumps effectively have the lowest plausible star-formation, the upper limits on the clumps CO(5-4) luminosity and inferred gas masses would be consistent with the SFE and gas fractions typically observed in MS galaxies, as illustrated by the dotted areas in Fig.~\ref{fig:SK}.
The measurement from the stacking analysis provides a tighter constraint on the gas deficit, but remains marginally consistent with the SFR-$M_{\rm gas}$ relation of MS galaxies, once the scatter of this relation is taken into account.

\subsection{Central CO(5-4) emission}

The fact that we detect CO(5-4) emission, which traces the dense molecular phase and hence the star formation, from the nuclear region in UDF6462 is in agreement with other studies showing that the absolute SFR profiles of star-forming, high-redshift galaxies are centrally peaked (although this is not the case for the sSFR).
For example, the analysis of the 3D-HST spectra in \citet{Nelson+16b} revealed $\Halpha$-based star formation surface density profiles peaking within the inner $\sim$1\,kpc across the entire MS at 0.7\,$<$\,$z$\,$<$\,1.5. 
Star formation coinciding with the central stellar mass concentration has also been found in sub-mm and radio observations of 1\,$<$\,$z$\,$<$\,3 galaxies \citep{Rujopakarn+16,Tadaki+17}. 
While the SFR and CO(5-4) peaks are coincident, the sSFR inferred from the SED fitting has formally a minimum over the nuclear region of UDF6462 (but is not as low as in quenched galaxies). As we already argued before, dust obscuration could affect the UV-based SFR at the centre of UDF6462 and the intrinsic sSFR profile could be steeper than measured.

What seems peculiar in UDF6462 is instead the presence of a CO(5-4) emission that is more centrally concentrated than the distribution of other star formation tracers (radio, UV or $H\alpha$) or of the dust continuum emission.
Differences between the size of star formation and of the CO emission have been already reported in the literature, although in these works trends opposite to that measured for UDF6462 were found: low-$J$ CO emission is typically more extended than star formation or dust continuum, and high-$J$ transitions trace quite closely the distribution of star formation \citep{Riechers+11,Hodge+12,Hodge+15,Spiker+15}. 
We present some possibilities that could explain our opposite results.

Recent numerical works \citep{Zolotov+15,Tacchella+16} have suggested that gas-rich, high-redshift galaxies experience dissipative gas inflow, a  so-called ``compaction" phase, which results in centrally peaked gas profiles and eventually leads to formation of a bulge with high central stellar surface density and to inside-out quenching of star formation, consistent with the observed structural properties of passive galaxies \citep[e.g.,][]{Wuyts+11,Genzel+14,Tacchella+15,Barro+16}.
An interesting possibility is that we observe a high concentration in the dense gas distribution (and a small gas content in the stellar clumps) because we caught UDF6462 during this particular phase of nuclear gas funnelling.
In this respect, we note that the stellar mass surface density\footnote{Derived from the \textsc{galfit} PSF-deconvolved best-fitting model  for the mass map.} within 1\,kpc for UDF6462 is currently $\Sigma_1$\,$\sim$10$^9\Msol$\,kpc$^{-2}$, i.e. below the typical central densities of $\sim$10$^{9.5-10}\Msol$\,kpc$^{-2}$ observed in quenched galaxies of a similar mass at z\,$\simeq$\,0  \citep{Fang+13,Tacchella+15}.
Taking at face value the gas masses inferred in Sections \ref{sec:GasFromCO} and \ref{sec:GasMassDyn} for the central  region and using the CO(5-4) intrinsic source geometry, we estimate a fraction of about 20-25\% of the total gas mass, i.e. ${\sim}6-8{\times}10^{9}\Msol$,  to be within the inner 1\,kpc. If all of this gas mass is converted into stars, the central mass density could reach a quenched-like value of $\sim$10$^{9.5}$ by $z\lesssim1$.

However, there are a few caveats to this interpretation. Inferring total gas content from CO(5-4) emission is subject to the uncertainties discussed in Section \ref{sec:GasFromCO} and it is hard to exclude that a more diffuse, less excited gas component is present outside the galaxy nucleus in UDF6462. The more extended distribution of the cold dust component would suggest that this is the case. Moreover, with typical compaction times of $\sim$0.5-1Gyr \citep{Zolotov+15} and time-scales of a few 100\,Myr for UV-based SFRs, the window for detecting a significant difference in the distributions of CO and UV emission is short.

If we take the extent of the dust distribution as evidence that the less excited, total gas reservoir is in fact more diffuse than the CO(5-4) component tracing star formation, an alternative possibility could be that we are observing a radial variation of the SFR/$M_{\rm gas}$ ratio, namely a peak in the SFE near the galaxy centre. The lower concentration of the UV- and $H\alpha$-based SFR could be a consequence of the significant dust obscuration over this region. Interaction-driven star formation could explain such a behaviour, but the origin of the enhanced SFE is unclear.

Finally, another potential explanation for our data could be a radial dependence of the $L^{\prime}_{\rm CO(5-4)}$--SFR relation, caused by changes in temperature, radiation field or stellar and gas density.
In nearby disc galaxies, high interstellar gas pressure leads to an increase of the dense gas fraction in galaxy centres. As this does not proportionally express itself in form of a higher SFR, variations of the ratio between the intensity of dense gas tracers and IR emission (e.g., $L^{\prime}_{\rm HCN}/L_{\rm IR}$) are observed between the inner and outer regions \citep{Usero+15,Bigiel+16}. The CO(5-4) peak in the centre of UDF6462 could therefore reflect a more efficient production of the dense gas phase within this region, compared to the less dense intra-clump medium.
Such variations may in general contribute to the scatter in the observed $L^{\prime}_{\rm CO(5-4)}$--$L_{IR}$ relation (as also evidenced by the fact that UDF6462 lies on the upper edge of this correlation in Figure \ref{fig:LirLCO}).
Feedback over the nuclear region could also affect the CO excitation. The presence of a strong AGN was excluded in Section \ref{sec:AGNcontribution}, but stellar feedback or past nuclear AGN activity could have resulted in highly excited gas over the galaxy centre.

Further observations, tracing different molecular transitions and species, would be required to discern among the scenarios discussed above.
As our observations have demonstrated, however, the detection of CO emission from resolved regions in high-$z$, MS galaxies requires several hours of integration on a single target even with ALMA, implying that studying   ISM at kiloparsec scales and within clumps in normal galaxies remains a challenge.

\section*{Acknowledgements}

We thank the anonymous Referee for their constructive comments. 
AC (ORCID 0000-0003-4578-514X) acknowledges support from the UK Science and Technology Facilities Council (STFC) consolidated grant ST/L000652/1. 
MTS acknowledges support from a Royal Society Leverhulme Trust Senior Research Fellowship (LT150041).
We acknowledge financial support from the Swiss National Science Foundation (AC, project PBEZP2\_137312), from the Agence Nationale de la Recherche  (AC, ElF, contract no. ANR- 12-JS05-0008-01) and from the E. C. through an ERC grant (AC and FB, StG-257720).
GEM acknowledges support from the ERC Consolidator Grant funding scheme (project ConTExt, grant number No. 648179) and a research grant (13160) from Villum Fonden. WR is supported by a CUniverse Grant (CUAASC) from Chulalongkorn University and JSPS KAKENHI grant no. JP15K17604.
This paper makes use of the following ALMA data: ADS/JAO.ALMA\#2013.1.01271.S. ALMA is a partnership of European Southern Observatory (ESO, representing its member states), NSF (USA) and NINS (Japan), together with NRC (Canada), NSC and ASIAA (Taiwan), and KASI (Republic of Korea), in cooperation with the Republic of Chile. The Joint ALMA Observatory is operated by ESO, AUI/NRAO and NAOJ.
The research leading to these results has received funding from the European Commission Seventh Framework Programme (FP/2007-2013) under grant agreement no. 283393 (RadioNet3).



\appendix

\section{Physical Properties Derived from Resolved \textit{HST} Imaging} \label{app:ClumpMeasurements}

\subsection{Comparison of resolved SFR maps}

As we have discussed in Section~\ref{sec:ResolvedMaps}, we derived several versions of the SFR and mass maps for UDF6462.  We use these maps to infer physical properties at the clump scale; hence, it is useful to quantify how sensitive are the inferred parameters to the method used for the calculation.

\begin{figure*}
\begin{center}
\includegraphics[width=0.8\textwidth,angle=90]{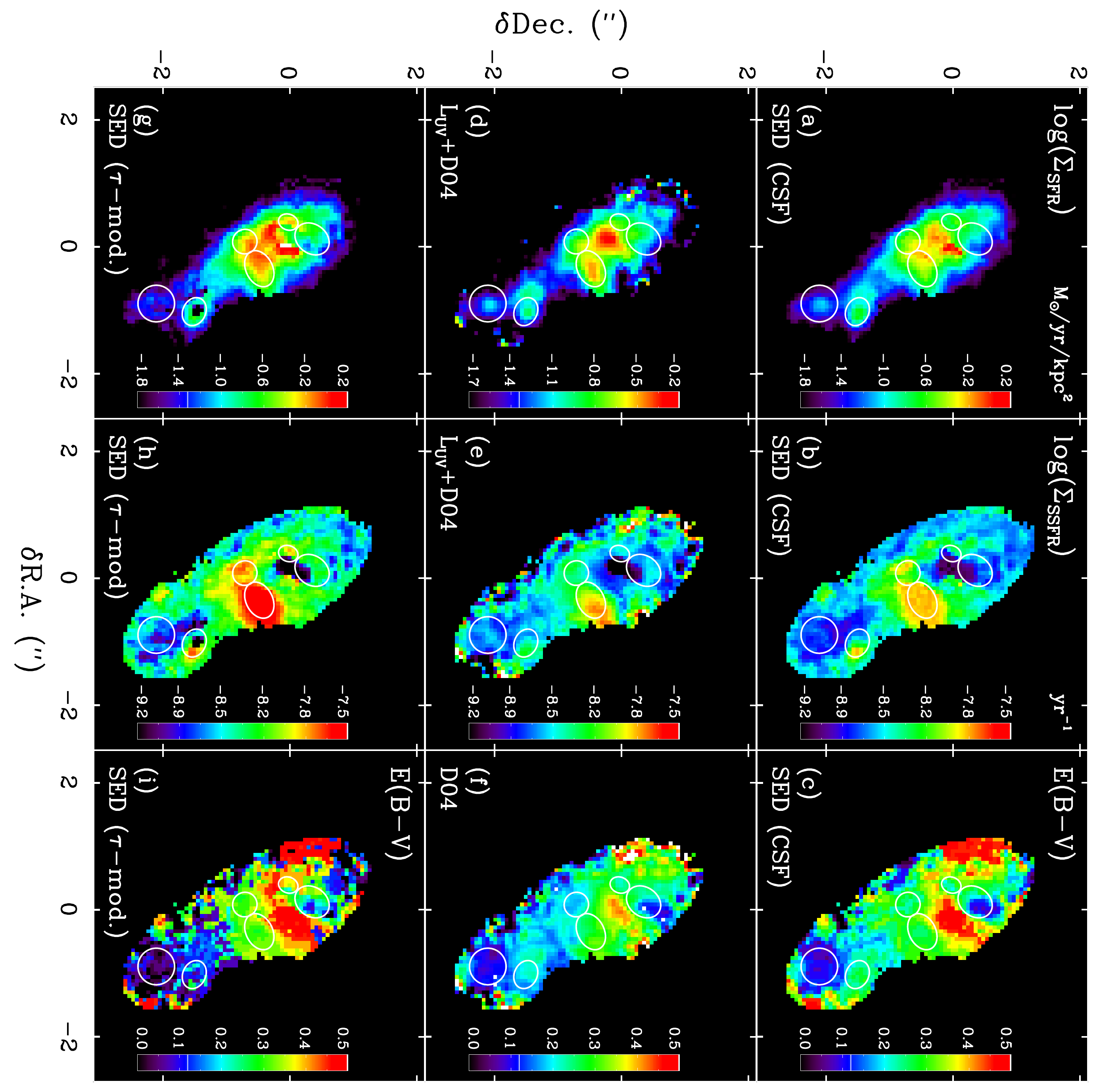}
\end{center}
\caption{\label{fig:UDF6462ResMaps_Appendix} SED-based SFR and extinction maps for UDF6462. (a) $-$(c): map of SFR density, sSFR and   $E(B-V)$ derived from SED fitting with CSF history templates;  (d)$-$(f):  map of SFR density, sSFR and $E(B-V)$ obtained applying the $L_{\rm UV}$-SFR and $E(B-V)$ versus $(b_{435}-z_{850})$ relations in \citet{Daddi+04}; (g)$-$(i): map of SFR density, sSFR and   $E(B-V)$ derived from SED fitting with delayed $\tau$-models.  Axis coordinates represent offsets with respect to the phase centre of the ALMA observations. White circles represent the position of the six clumps as in Fig.~\ref{fig:UDF6462Stamps}.
}
\end{figure*}

While for the mass maps we obtain very similar results regardless of the templates employed for the SED fitting ($\tau$ versus CSF), not surprisingly larger variations are found for the SFR maps.
We show in figure \ref{fig:UDF6462ResMaps_Appendix} all the available SFR maps and the associated extinction and sSFR maps.
Qualitatively, the different maps provide similar global trends placing most of the SFR  over the nuclear region, where also the dust extinction is peaking, with some activity extending towards the two Southern clumps (labelled with numbers 5 and 6 in Fig.~\ref{fig:UDF6462Stamps}). 
As mentioned in Section~\ref{sec:ResolvedMaps}, the map derived using $\tau$-models is the noisier one, displaying step-like variations of the SFH between neighbouring regions (see e.g. clump 5). For this reason, this map is not considered further for deriving clump properties.

\begin{figure}
\begin{center}
\includegraphics[width=0.5\textwidth]{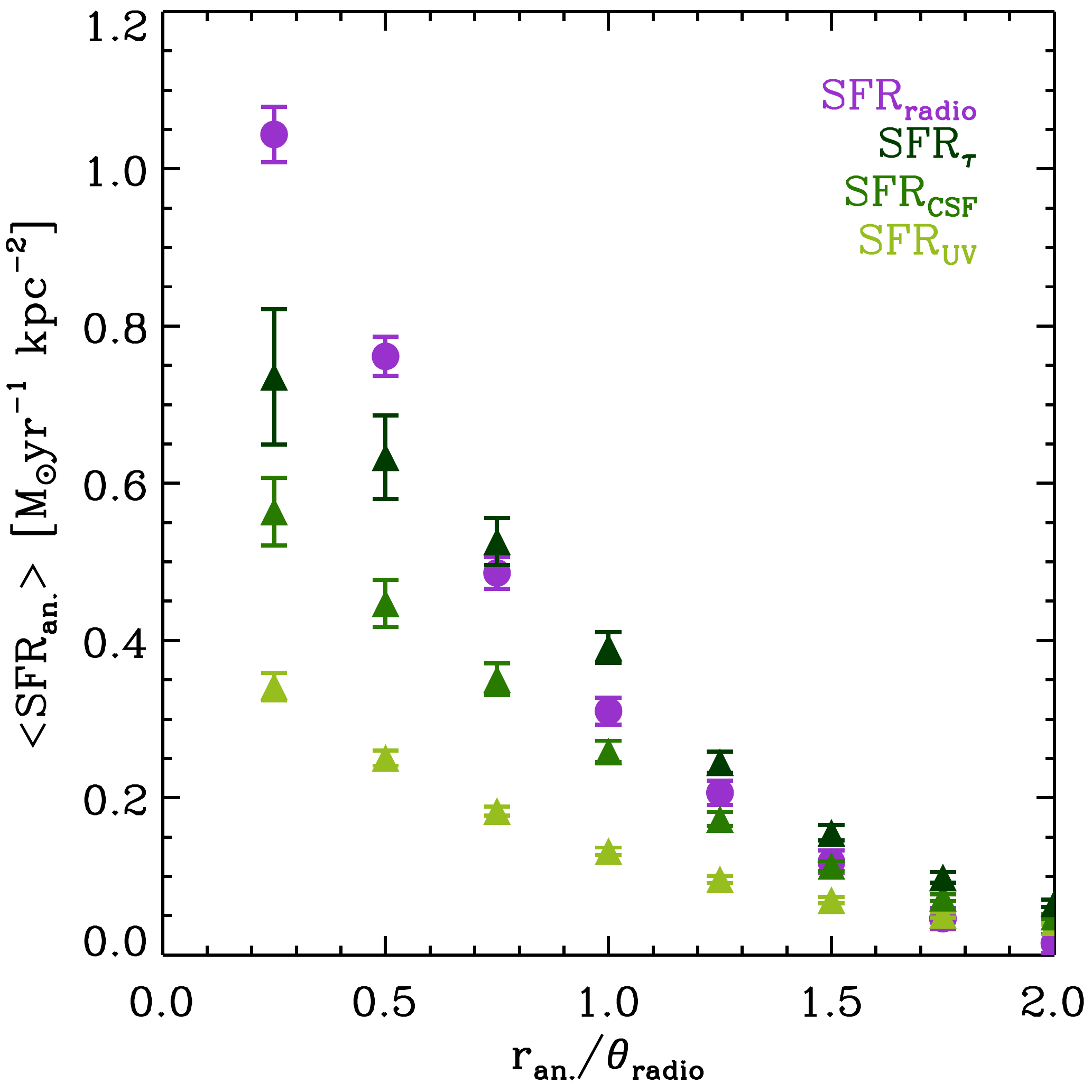}  
\caption{SFR density profile of UDF6462 derived from either the VLA radio map (purple symbols) or from the different versions of the SED- or UV-based SFR maps (green symbols, see figure legend). The average SFR is calculated in running annuli of width 0.5$\times \theta_{\rm radio}$, where $\theta_{\rm radio}$ is the FWHM size of the radio beam. The \textsc{HST} SFR maps have been degraded to the VLA resolution for this comparison.}
\label{fig:App:SFRProfiles}
\end{center}
\end{figure}

Quantitatively, however,  the SFR can vary substantially at the pixel-to-pixel scale from one map to the other. 
The largest differences are observed in the nuclear region as a consequence of the significant dust extinction in this area. To understand whether the maps derived from the optical/NIR \textit{HST} images can robustly recover the obscured SFR, we compare in Fig.~\ref{fig:App:SFRProfiles} the SFR density profile from the pixel-by-pixel SED fitting with that inferred from the VLA, dust-free radio map. 
Here we assume that the radio emission is dominated by star formation with no AGN contribution, which we argued to likely be the case in Section~\ref{sec:AGNcontribution}  and we convert the observed radio fluxes profiles to SFR following the same procedure applied for the integrated galaxy in Section \ref{sec:SFRMassInt}.  To perform this comparison, we convolved the SFR maps, natively at the \textit{HST} resolution (0$\farcs$15), with a kernel matching the radio beam ($\theta_{\rm radio}$=0$\farcs3\times 0\farcs6$).

Interestingly, as shown in Fig.~\ref{fig:App:SFRProfiles}, all SED maps underestimate the SFR with respect to the radio within a radius $\sim \theta_{\rm radio}$. This bias is most pronounced for the SFR map derived from the UV luminosity (light green symbols) which presents SFR that are a factor $\sim$\,2.5 lower than the radio. However, also in the CSF (and $\tau$) map the nuclear SFR is underestimated, by a factor $\sim$\,1.5, in spite of the fact that the sum of the pixels in this map is comparable to the galaxy total SFR.  
At radii $\gtrsim \theta_{\rm radio}$, where most of the clumps are located and the extinction is low, the UV and CSF SFR maps are instead in better agreement with the radio data and can thus be used to infer the clump properties.

\subsection{Clumps at different wavelengths and estimates of star formation rates and masses}  \label{app:ClumpsProps}

\begin{figure}
\begin{center}
\includegraphics[width=0.5\textwidth]{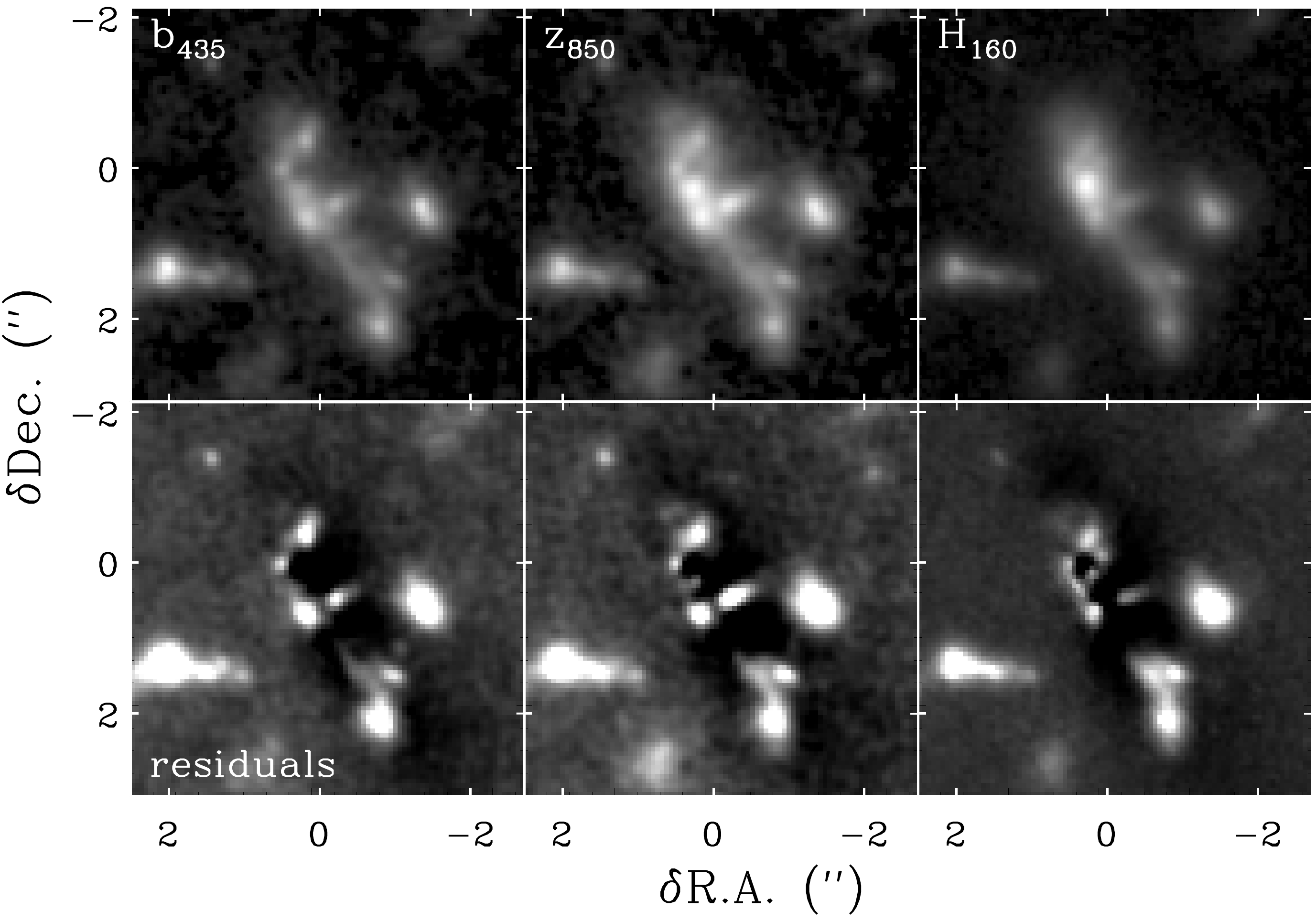}  
\caption{\emph{Top:} original XDF $b_{435}$, $z_{850}$ and $H_{160}$ images of UDF6462. \emph{Bottom}: residual images after subtraction of the \textsc{galfit} best-fitting model of the galactic diffuse component (see Section~\ref{sec:ClumpsExtraction}). The six clumps identified in UDF6462 are visible in all three bands.}
\label{fig:AppResiduals}
\end{center}
\end{figure}

As described in Section~\ref{sec:ClumpsExtraction} our identification of the clumps in UDF6462 is not strongly wavelength dependent. We illustrate this in Fig.~\ref{fig:AppResiduals} where we present the residual images used to identify the clumps in the $b_{435}$, $z_{850}$ and $H_{160}$.
The six main clumps are clearly visible in all three images and the identification is not affected by strong colour gradients.  For easier comparison with previous works on this galaxy and given that clumps have a higher SNR in this band, we used the $z_{850}$ filter as our extraction image.

In determining the photometry and physical properties of the clumps, one of the major uncertainties is whether and how to account for the underlying galactic diffuse component. 
Here we followed a number of different approaches which bracket several assumptions for this contribution.

The first method we employed to estimate clump SFRs and masses was to perform SED fitting to the $UV_{275}$ to $H_{F160}$ integrated photometry extracted over the petrosian apertures defined by the white circles in Fig.~\ref{fig:UDF6462Stamps}. 
Clump magnitudes were calculated either by simply summing up the fluxes in all pixels within the apertures -- namely, assuming that all the emission is associated to the clumps only --  or by subtracting a diffuse component to these fluxes.
To minimize model dependencies, we decided to quantify the diffuse component by simply taking the average surface brightness of pixels that are not in the clump regions, as done in \citet{Guo+12}. 
Given that UDF6462 displays a North-South colour gradient (see Fig.~\ref{fig:UDF6462Stamps}), we calculated the average surface brightness for the four Northern clumps and the two Southern clumps separately. We then subtracted these estimates, multiplied by the area of the pixels, to the total flux calculated over the clumps. 

Another method was to sum up the pixels  in the resolved SFR or mass maps over each clump area. This approach also assumes no contribution from the underlying galactic diffuse component and hence provides an upper limit to the actual SFR or mass in clumps.
 
Finally, we also performed a \textsc{galfit} decomposition of the mass maps and SFR into the galactic and clump components. Such a seven components fit (galaxy+six clumps) can quickly become very degenerate if left unconstrained. 
To minimize the number of free parameters used in the fit, we thus applied the following procedure.  
We first carried out a single S\'ersic fit to the mass map masking out the clump regions and allowing some degree of asymmetry (first azimuthal Fourier mode in \textsc{galfit}). Clumps contribute only a minor fraction of the galaxy total mass and the mass map is therefore the closest estimate of the smooth underlying component that is available.  
We then performed the galaxy+clumps decomposition keeping the basic structural properties of the underlying smooth galactic component (position angle, ellipticity and lopsidedness) fixed to those obtained from the single S\'ersic fit on the mass map. When fitting the mass map, all other parameters, except the total mass, were also kept fixed to the single component solution. For the SFR maps, we instead allowed the half-light radius and index $n$ to vary as well,  but used the results from the single S\'ersic model for the diffuse component in the mass map as initial guesses. The clumps were modelled as elliptical Gaussians (with the exception of Clump \#3 which is best described by a point source). We used the best-fitting Gaussian parameters from the fit to the residual $b_{850}$-band image described in Section \ref{sec:ClumpProp} as initial guesses and constrained the Gaussians FWHM to with $\pm$50\% of the $b_{850}$ FWHM.

The results of these calculations are shown in Fig.~\ref{fig:ClumpsSFR}. For comparison, we also show in that figure the mass and SFR derived by \citet{Guo+12} for the four clumps that are in common to their and our sample.

\section{Other Sources in the ALMA Field and Astrometry Offset with Respect to \textit{HST} imaging}\label{app:AstrometryOff}

\begin{figure*}
\begin{center}
\includegraphics[width=0.95\textwidth]{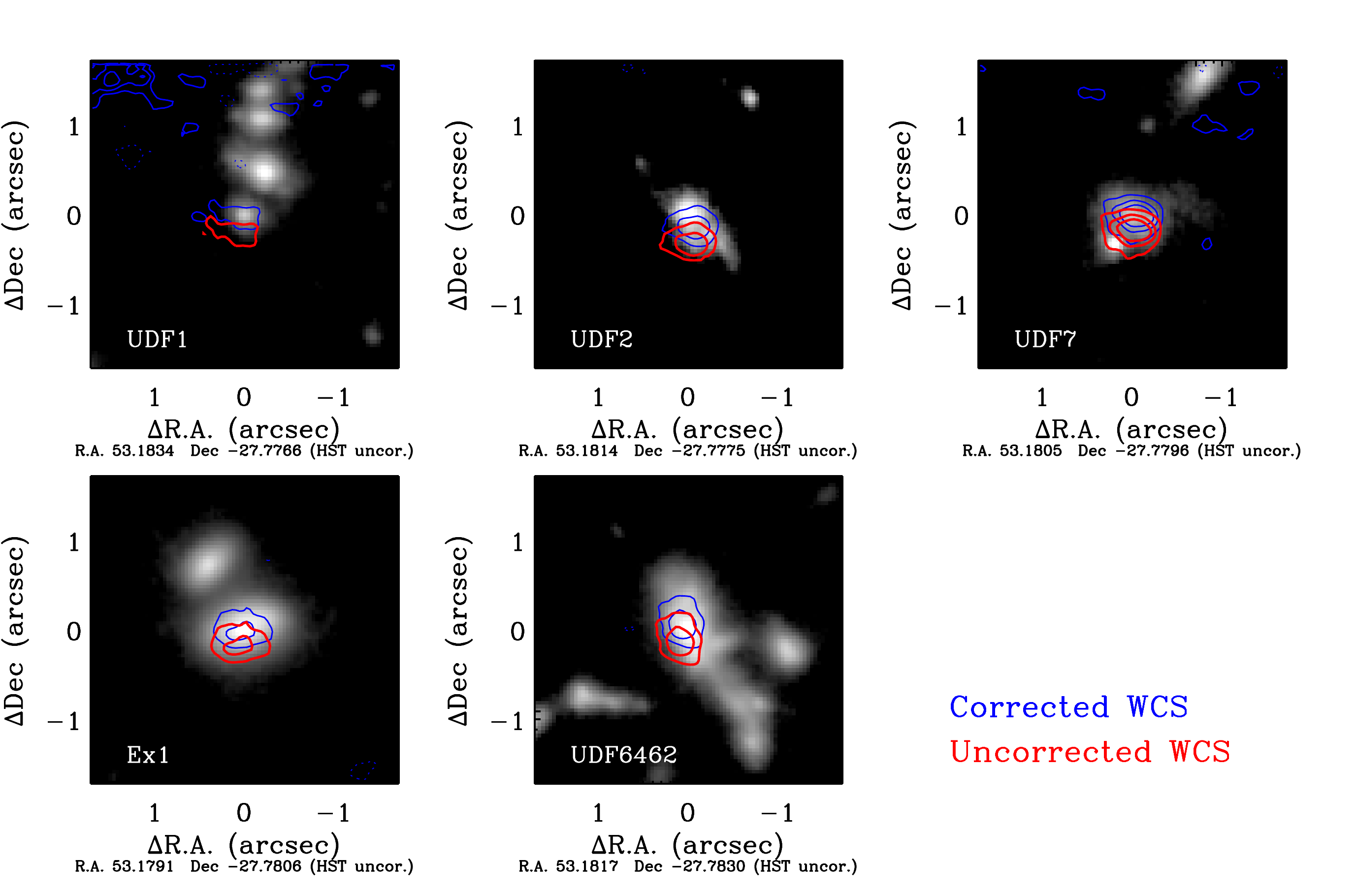}  
\caption{XDF $H_{160}$ images for UDF6462 (top left) and other four galaxies detected at the $>3\sigma$ level in the ALMA continuum image. 
The \textit{HST} images have been corrected for the astrometry offset that we found with respect to the ALMA data (correction applied: $\Delta$RA\,=$0\farcs06$ and $\Delta$Dec\,=$-0\farcs272$). The solid blue lines show 3$\sigma$, 5$\sigma$, and 9$\sigma$ contours in the CO(5-4) emission for UDF6462 and in the continuum for the other four galaxies (-3$\sigma$ levels shown as dotted lines). After applying the astrometric correction, we find a good agreement between the position of the \textit{HST} and ALMA sources.
The location of the ALMA contours prior to the astrometric correction is shown with the red lines.
 Coordinates beneath the axis labels correspond to the original uncorrected \textit{HST} positions.
The higher noise in the ALMA contours for UDF1 and UDF7 reflect the fact that these galaxies are close to the primary beam edge.}\label{fig:ALMAoffset}
\end{center}
\end{figure*}
Together with UDF6462, four other galaxies are clearly detected at a $>3\sigma$ level in our deep, 1300\,$\mu$\,m continuum image. The XDF $H$-band cutouts of these sources are shown in Fig.~\ref{fig:ALMAoffset} and a summary of their properties is given in Table \ref{tab:contSources}.
Three of these galaxies were also studied in the work of \citet{Dunlop+17} and \citet{Rujopakarn+16}, enabling us to independently test our flux and size measurement techniques. We find a  good agreement between our estimates and those previously published  by these authors, although we note that UDF1 (and to a smaller extent UDF7) lies at the edge of the ALMA field of view, making it subject to noise effects and large primary beam correction. 
A fourth galaxy (labelled Ex1) was not previously detected and we report its 1300\,$\mu$m continuum flux for the first time here.
Near the edge of the field of view of our ALMA observation falls another galaxy (UDF16) from the sample of \citet{Dunlop+17}. 
At the reduced sensitivity over this region of the ALMA map,  this source is not detected in our data.

We utilized these four galaxies, plus UDF6462, to correct the astrometric offset between ALMA and \textit{HST} imaging, which we discovered when analysing the ALMA data.
Overlaid on the $H$-band images in Fig.~\ref{fig:ALMAoffset}  are the ALMA contours of either the CO(5-4) emission, in the case of UDF6462, or the continuum emission for the other four galaxies. A systematic shift of the \textit{HST} centroids with respect to the ALMA data is evident in all cases. This shift is mostly in the North-South direction.  We quantified the magnitude of the offset as the median difference between the peak coordinates in the $H_{160}$-band and ALMA images for these four galaxies. We found that the \textit{HST} positions are shifted by $\Delta$RA\,=$-0\farcs$06  and $\Delta$Dec\,=$+0\farcs272$ with respect to ALMA. 
We estimate positional uncertainties for ALMA sources to be given by $\theta_{\rm beam}/(SNR)/\sqrt{4\log 2}$\,$\simeq$\,$\theta_{\rm beam}$/(2$\times$SNR) \citep{Ball1975,Condon_1997}, namely approximately 100, 70 and 40\,mas for signal-to-noise ratios SNR\,=\,3, 5 and 10, respectively. The observed shift in declination is therefore larger than the ALMA positional uncertainty. 

Inconsistency between \textsc{HST} and ALMA coordinate systems for the GOODS-S field are a known issue and were already reported in the literature. In their analysis of ALMA 1.3mm and VLA 5cm observations, \citet{Rujopakarn+16}
found offsets between ALMA/VLA and \textsc{HST} astrometry over the HUDF area to be $\Delta$RA\,\,=\,$-0\farcs076$ and 
$\Delta$Dec\,=$+0\farcs279$. Similar offsets were also found in \citet{Barro+16} and  \citet{Dunlop+17}.
These previous estimates are in very good agreement with those calculated from our own data.
Hence we corrected the \textsc{HST} world coordinate system with the offsets we have derived here (note that opposite corrections to those provided above have to be applied to the \textit{HST} imaging) and we refer only to the corrected images in this work.

\begin{table*}
	\centering
	\caption{Properties of the four galaxies detected in our ALMA continuum image, together with UDF6462. Column 1 is the galaxy ID.  Galaxies in rows 1 to 4 have the same ID as in \citet{Dunlop+17} and the galaxy labelled as \emph{Ex1} was not detected by these authors. Columns 2 to 5 are source the best-fitting coordinates, FWHM and $1300 \mu m$ fluxes obtained by Gaussian fitting in the $uv$-plane. Celestial coordinates correspond to the ALMA astrometric system. Columns 6 and 7 are the sizes and fluxes for these sources published in \citet{Rujopakarn+16} and \citet{Dunlop+17}, respectively. Galaxy UDF16 is not detected in our data, the $3\sigma$  upper limit for its flux in column 5 is derived by fitting a Gaussian  with parameter fixed to those provided in \citet{Rujopakarn+16}. }
	\begin{tabular}{ccccccc} 
		\hline \hline
		ID & RA$_{\rm ALMA}$ & Dec.$_{\rm ALMA}$  & $\theta_{\rm major} \times \theta_{\rm minor}$ & S$_{1300 \mu m}$  & $\theta_{\rm major} \times \theta_{\rm minor}$[R16] & S$_{1300 \mu m}$[D17]\\
	          & (deg) & (deg) & ($^{\prime \prime}$) & $(\mu$\,Jy) & ($^{\prime \prime}$) & $(\mu$\,Jy) \\
		\hline \\
		UDF1  & 53.183502   &  -27.776686   &  0.51$\pm$0.14\,$\times$\,0.20$\pm$0.08   &1063$\pm$132 &  
		                                                                 0.39$\pm$0.04\,$\times$\,0.33$\pm$0.04   & 920$\pm$49 \\
                UDF2  & 53.181389   &   -27.777584   &  0.47$\pm$0.05\,$\times$\,0.28$\pm$0.05   &   875$\pm$47  &
                		                                                         0.53$\pm$0.06\,$\times$\,0.45$\pm$0.05   & 996$\pm$87 \\
                UDF7  & 53.180531   &   -27.779709   &  0.34$\pm$0.10\,$\times$\,0.10$\pm$0.06   &   234$\pm$12 & 
                                		                                        $<$0.24\,$\times$\,$<$0.13   & 225$\pm$13 \\
                UDF16  & 53.180531   &   -27.779709   &   --   &  $<$164  & 
                                		                                        $<$0.24\,$\times$\,$<$0.15   & 155$\pm$44 \\
                Ex1      & 53.179134   &   -27.780613   &  0.36$\pm$0.04\,$\times$\,0.29$\pm$0.03   &   315 $\pm$13  & -- & --    \\
		\hline
	\end{tabular} \label{tab:contSources}
\end{table*}

\section{Tests on Simulated Images}\label{app:Sim}

\begin{figure*}
\begin{center}
\begin{tabular}{lc}
\includegraphics[width=0.375\textwidth]{fC2a.pdf}   & \includegraphics[width=0.335\textwidth]{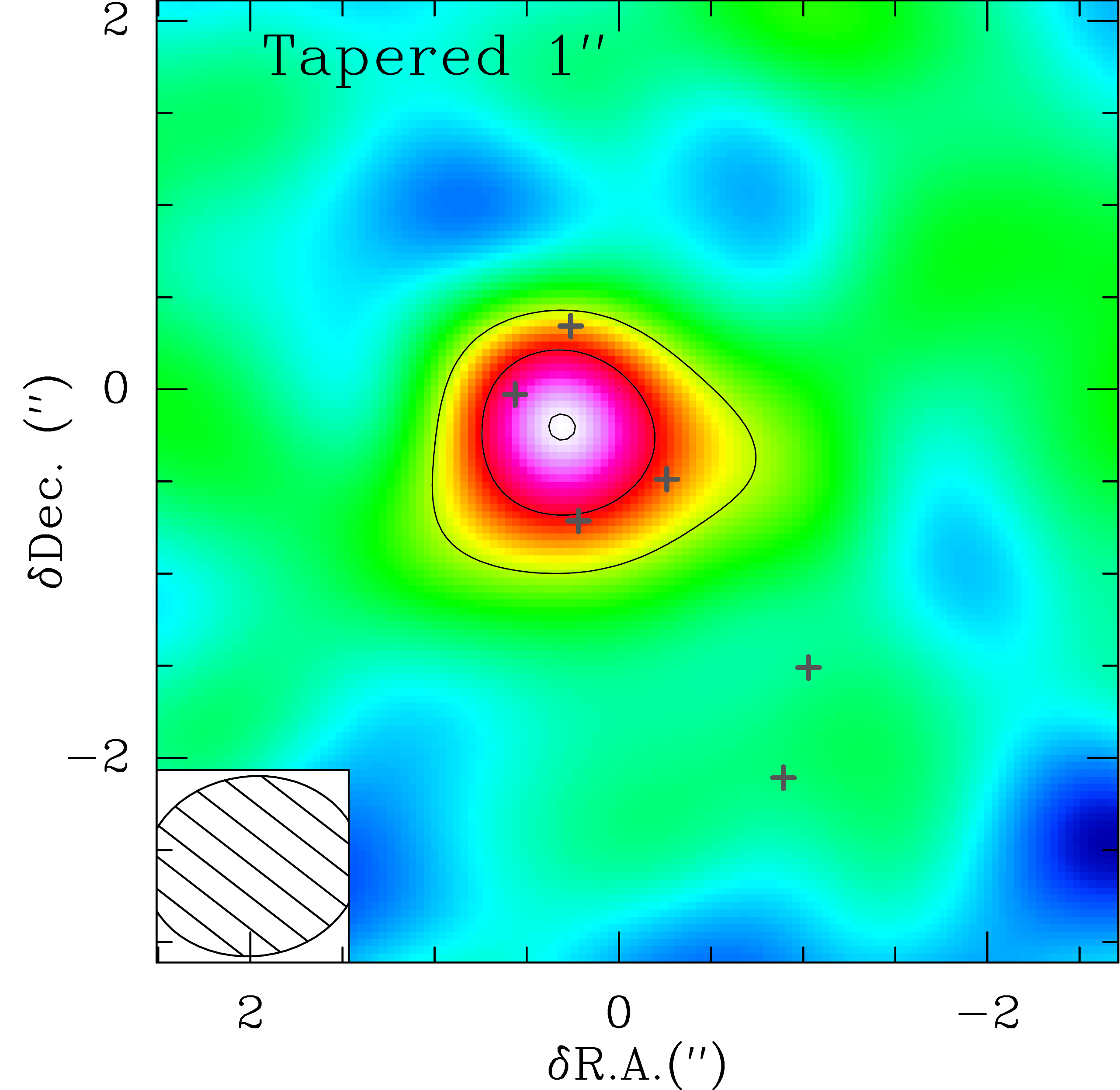} \\
\includegraphics[width=0.335\textwidth]{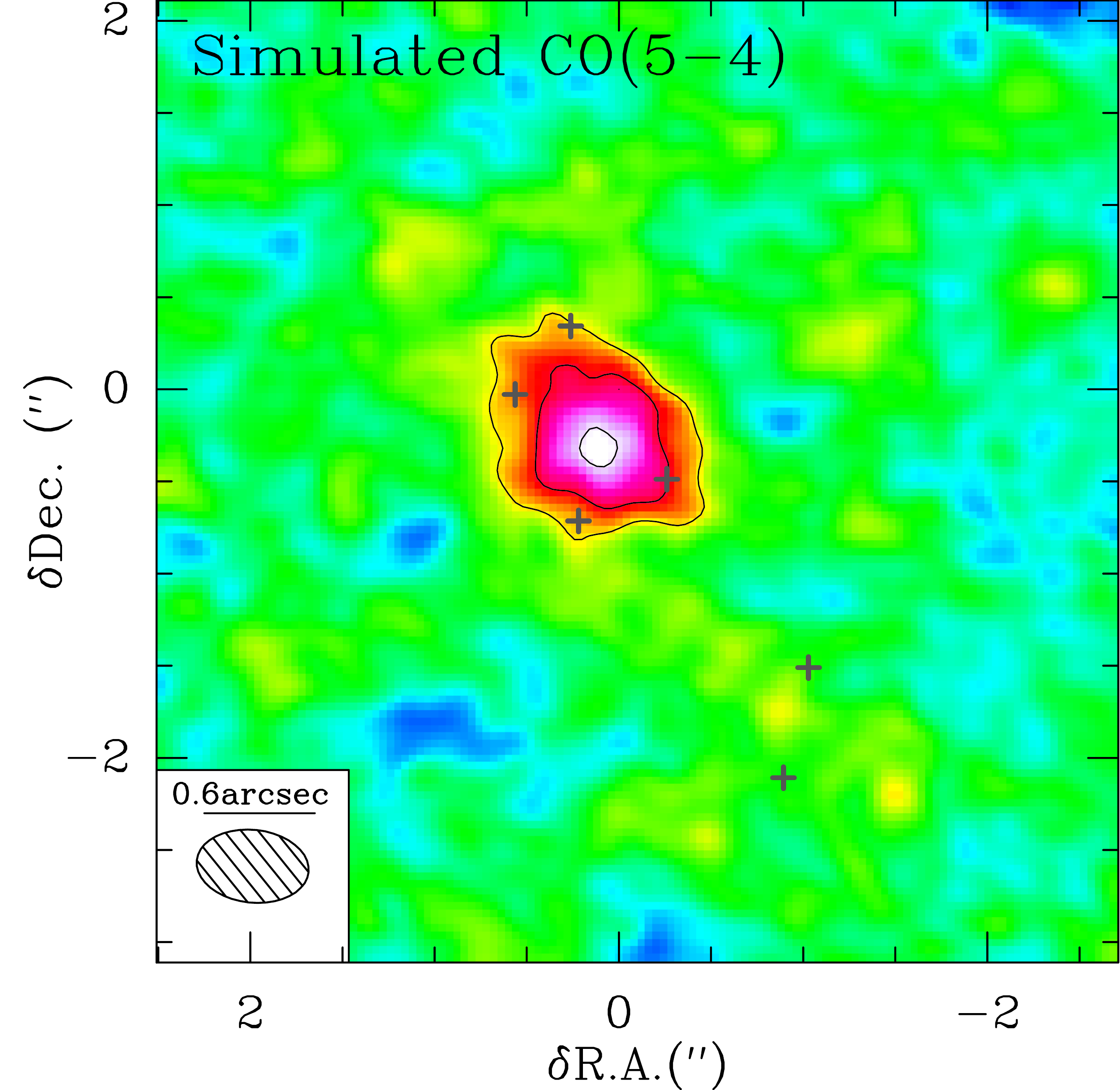}   & \includegraphics[width=0.335\textwidth]{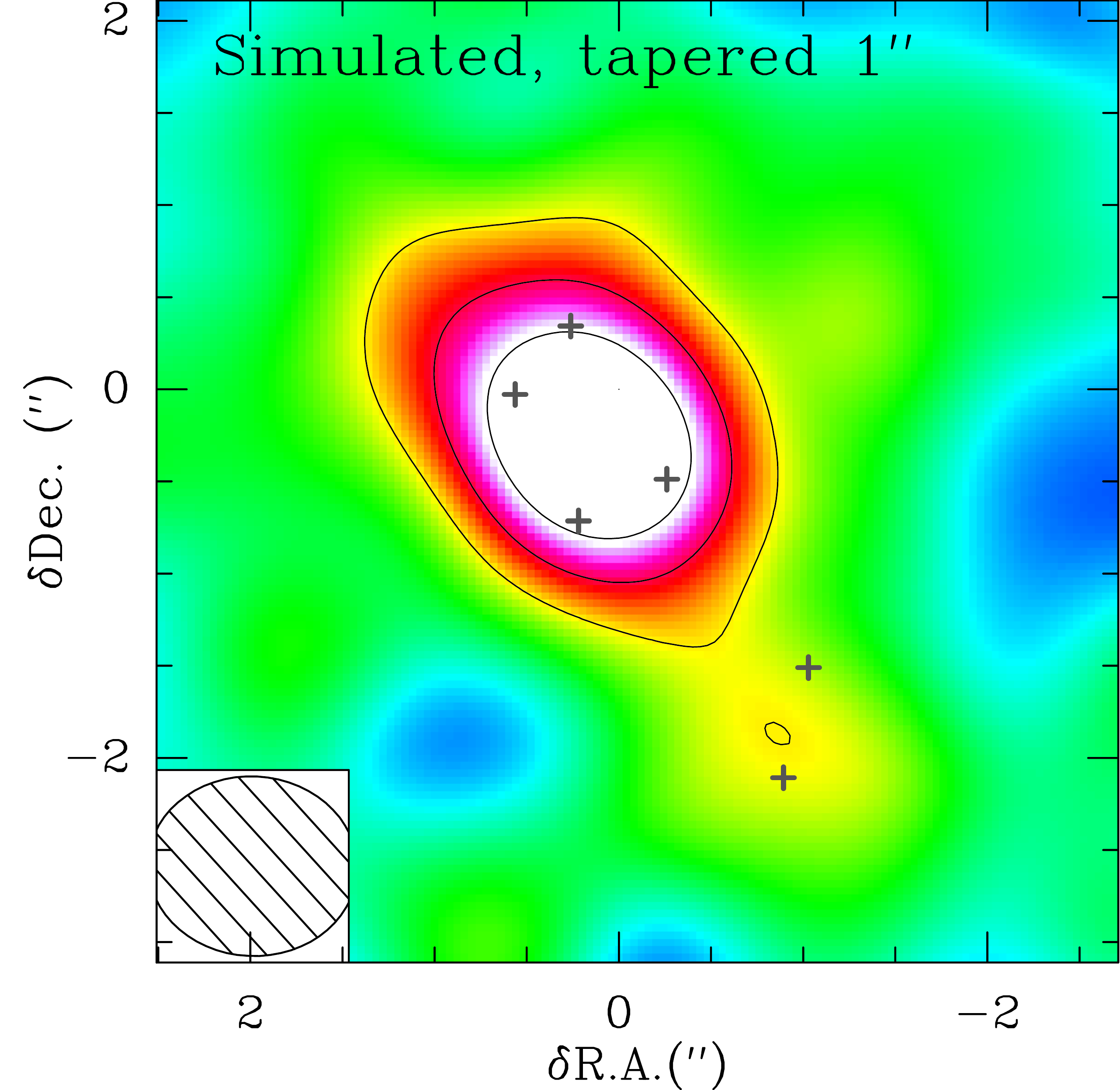} \\
\end{tabular}
\caption{Comparison of the real CO(5-4) observations of UDF6462 (top row) with the simulated ALMA observations generated from the UV-based SFR map (bottom row).
On the left we show the clean maps obtained at the native 0$\farcs$6$\times0\farcs4$ resolution of the ALMA observation (top left panel is identical to Fig.~\ref{fig:COlineInt}), on the right after tapering on a scale of 1$^{\prime \prime}$. Identical bitmap scaling is used in the real and simulated images. Contours denote the [3, 6, 9]$\times$rms levels. Grey crosses highlight the position of the clumps.}
\label{fig:App:FakeObs}
\end{center} 
\end{figure*}

We describe here the simulated ALMA observations utilized to test the robustness of the CO(5-4) size measurements.
We started by generating an input model for the CO(5-4) emission from the \textit{HST} SFR map. We show here simulations based on the SFR map obtained from the dust-corrected $L_{\rm UV}$, but similar results are obtained if using the SFR from the SED fitting with CSF models.  As we discussed in Section~\ref{sec:COSFRdist}, a direct conversion of the SFR into CO(5-4) luminosity results in too low fluxes, likely as a consequence of underestimation of nuclear dust obscuration. For this reason, we renormalized the input SFR map such to match the observed CO(5-4) line flux within the $0\farcs5\times 0\farcs2$ Gaussian region over which we detect the line emission, accounting for \textit{HST} PSF blurring.
(Note that this effectively translates into a simple division of the SFR map by the flux within that area; hence, the pixel relative fluxes and radial distribution are conserved.)
We then employed the \textsf{simobserve} task within \textsc{casa} to simulate ALMA observations. The relevant task parameters were set to reproduce exactly the date and time of observation, antennas configuration, the total integration time and atmospheric conditions of the real ALMA data.
During the simulation, we employed a channel width of 0.33GHz, i.e., corresponding to the $\sim$430km/s velocity range over which we detect significant line emission (grey shaded area in right-hand panel in Fig.~\ref{fig:COlineInt}).

We show in Fig.~\ref{fig:App:FakeObs} the obtained artificial ALMA maps. Globally, the simulated observations reproduce the real data in terms of SNR and spatial distribution. 
We fit the simulated $uv$-plane with a Gaussian source, using the \textsf{UV\_FIT} task as done for the real observation and measured a size of FWHM=(1$\farcs$1\,$\pm$\,0$\farcs$1)\,$\times$\,(0$\farcs$7$\pm$\,0\farcs1), i.e., we recovered the intrinsic model size. Tapered versions of both the real and simulated images on a scale of 1$^{\prime \prime}$ are shown in the right-hand panels of Fig.~\ref{fig:App:FakeObs}. While tapering on the simulated data reveals the extended emission in the North-South direction and increases SNR, in the real ALMA observations this is not the case and the SNR is not significantly improved.
The findings above therefore suggest that emission distributed on scales of 1$^{\prime \prime}$, if present, should be recoverable at the SNR of our observations and that the small CO size compared to the SFR/radio map is likely intrinsic.   


\bsp	
\label{lastpage}
\end{document}